\documentclass[11pt,a4paper]{article}
\usepackage{jheppub,amsmath, amsthm, amssymb,slashed,url,bbold,xcolor,upgreek}
\usepackage{xparse}
\usepackage{graphicx}

\graphicspath{{./FIGS/}}

\usepackage{boldline}
\usepackage{bm}
\usepackage{amssymb,amsfonts}
 
\usepackage{mathtools}
\DeclareMathOperator\arctanh{arctanh}

\DeclareDocumentCommand\dd{ E{^}{{}} m }{
  \mathinner{\mathrm{d}^{#1}#2}
}

\DeclareDocumentCommand\DD{ E{^}{{}} m }{
  \mathinner{\mathcal{D}^{#1}#2}
}

\DeclareDocumentCommand\Op{ m m}{
  \mathcal{O}_{#1,#2}
}

\DeclarePairedDelimiter\eval{.}{\rvert}
\DeclarePairedDelimiter\ev{\langle}{\rangle}
\DeclarePairedDelimiter\pqty()
\DeclarePairedDelimiter\bqty[]

\DeclareMathOperator{\Tr}{Tr}
\newcommand*{\hypF}{\prescript{}{2}{F}_{1}}

\RequirePackage{lmodern}
\usepackage[T1]{fontenc}

\usepackage{acronym}
\acrodef{eft}[\textsc{eft}]{effective field theory}
\acrodef{eom}[\textsc{eom}]{equation of motion}
\acrodef{ir}[\textsc{ir}]{infrared}
\acrodef{uv}[\textsc{uv}]{ultraviolet}
\acrodef{gny}[\textsc{gny}]{Gross--Neveu--Yukawa}

\usepackage{etoolbox}%
    \makeatletter
    \patchcmd{\maketitle}{\@fpheader}{}{}{}
    \makeatother

\def\XXint#1#2#3{{\setbox0=\hbox{$#1{#2#3}{\int}$}
     \vcenter{\hbox{$#2#3$}}\kern-.5\wd0}}

\def\nab{\overrightarrow{\nabla}}
\def\nabl{\overleftarrow{\nabla}}

\def\OMIT#1{{}}

\def\llra{{\relbar\joinrel\longrightarrow}}
\def\mapright#1{{\smash{\mathop{\llra}\limits_{#1}}}}

\newcommand{\beq}{\begin{equation}}
\newcommand{\eeq}{\end{equation}}
\newcommand{\beqa}{\begin{eqnarray}}
\newcommand{\eeqa}{\end{eqnarray}}

\newcommand{\kf}{k_{\rm F}}

\def\kf{k_{\rm F}}

\begin{document}

\title{Exact evaluation of large-charge correlation functions in non-relativistic conformal field theory}
\vskip 0.5cm
\author[a,c]{Silas R.~Beane,}
\author[b,a,d]{Domenico Orlando,}
\author[b,a]{and Susanne Reffert} 

\affiliation[a]{Albert Einstein Center for Fundamental Physics,
  Institut f\"ur Theoretische Physik,\\ Universit\"at Bern,
  Sidlerstrasse 5, CH-3012 Bern, Switzerland}
\affiliation[b]{Yukawa Institute for Theoretical Physics, Kyoto University, Kyoto 650-0047, Japan}
\affiliation[c]{Department of Physics, University of Washington,
  Seattle, WA 98195}
\affiliation[d]{INFN sezione di Torino.
via Pietro Giuria 1, 10125 Torino, Italy}

\subheader{YITP-24-48,NT@UW-24-05}
\vphantom{} \vskip 1.4cm

\abstract{The large-charge master field which generates all $n$-point
  correlation functions with an insertion of large charge Q in
  non-relativistic conformal field theory is obtained. This field is
  used to compute Schr\"odinger-invariant $n$-point correlation
  functions of large-charge operators via a direct evaluation of the
  path integral. Conformal dimensions are found to agree with
  calculations based on the state-operator correspondence. The master
  field solution exhibits an emergent harmonic trap whose frequency is
  a function of the Euclidean time. The large-charge effective action
  with operator insertions describes a droplet of superfluid matter
  whose spatial size scales with the time separation of sources. The
  solution is used to compute Schr\"odinger symmetry breaking
  corrections in the large-charge effective field theory (EFT) due to
  a finite scattering length in the fundamental theory of fermions
  near unitarity. The scaling of these effects in the large-charge
  power counting scheme is established, and the size of the effects is
  quantified using input from quantum Monte Carlo simulations of the
  near-unitary gas, as well as from the large-$N$ expansion at large charge.}

\maketitle

\section{Introduction}
\label{sec:intro}

In physical systems with many-body complexity, theoretical
progress relies on finding small parameters which enable a
perturbative expansion of physical observables. These small parameters
may arise from ratios of characteristic physical scales as in EFT
constructions, as well as from the large dimensionality of group
representations or large particle number. For momentum transfers that
are small compared to the pion mass, nuclear physics can be described
efficiently by an EFT of contact
operators~\cite{vanKolck:1998bw,Kaplan:1998tg,Kaplan:1998we}. As
s-wave scattering lengths in the nucleon-nucleon system are very large
compared to the range of the interaction, this EFT is
non-perturbative in the scattering length with higher-order terms in
the effective range expansion treated as perturbative insertions in
the EFT. If one considers a system of neutrons only, the
neutron-neutron scattering length is near infinite as compared to the
pion Compton wavelength, and the leading order in the nuclear EFT
is very near unitarity, where the theory is at a fixed point of the
renormalization group (RG) and is therefore described by a non-relativistic
conformal field theory, constrained by Schr\"odinger symmetry.  While
this EFT is significantly constrained by its proximity to
unitarity~\cite{Mehen:1999qs,Mehen:1999nd}, the machinery of conformal
field theory has not found many useful applications in this EFT
of nucleon contact operators.

This situation has recently changed with work of Hammer and
Son~\cite{Hammer:2021zxb} (See also
Ref.~\cite{Schaefer:2021swu,Braaten:2023acw,Chowdhury:2023ahp}) which
has shown that in special nuclear reactions with some number of
low-energy neutrons in the final state, non-relativistic conformal
field theory technology can be used to compute the part of the nuclear
reaction cross sections which account for neutron final-state
interactions. This EFT for nuclear reactions has been dubbed
``unnuclear'' physics (in analogy with the related ``unparticle''
physics of Ref~\cite{Georgi:2007ek}), as the object that describes the
propagating nucleus does not have a particle interpretation.  In the
EFT description, the leading order is a non-relativistic conformal
field theory, whose correlation functions are constrained by
Schr\"odinger
symmetry~\cite{Hagen:1972pd,Niederer:1972zz,Henkel:1993sg,Mehen:1999nd,Henkel:2003pu}. A
powerful tool for such theories is the state-operator
correspondence~\cite{Nishida:2007pj}, which allows the simple
computation of the conformal dimensions of operators that appear in
the correlation functions by considering the energy eigenspectrum of
the system in a harmonic trap. Schr\"odinger symmetry breaking effects
due to a finite scattering length and effective range can be included
in conformal perturbation theory~\cite{Chowdhury:2023ahp}.  This
method has now been applied to nuclear reactions with up to three
neutrons as one of the reaction products. However, as pointed out in
Ref.~\cite{Chowdhury:2023ahp}, the analysis of final states with more
than three neutrons is constrained by the complexity of the few-body
wavefunctions, and, for systems with many neutrons, it may be
profitable to consider the large-charge
expansion~\cite{Favrod:2018xov,Kravec:2018qnu} for non-relativistic
systems.

Based on earlier work with relativistic superfluids (for a review, see
Ref.~\cite{Hellerman:2015nra,Gaume:2020bmp}), the large-charge
expansion is formulated in the superfluid phase where the fundamental
degree of freedom is the Goldstone boson of spontaneously broken particle number~\cite{Greiter:1989qb,Son:2005rv}.  
Working in sectors of large global charge has been shown to lead to important simplifications and allows the computation of the conformal data in strongly-coupled relativistic CFTs~\cite{Hellerman:2015nra,Gaume:2020bmp}. Also in the non-relativistic case, the conformal
dimensions of charged operators have been computed to several
non-trivial orders, making use of the state-operator
correspondence~\cite{Favrod:2018xov,Kravec:2019djc,Hellerman:2020eff,Pellizzani:2021hzx,Hellerman:2021qzz,Hellerman:2023myh}.
A primary goal of this paper is to verify results based on the
state-operator correspondence via a direct evaluation of the path
integral, with no input from the state-operator correspondence. It
turns out to be straightforward to find the large-charge ``master
field'' solution of the Goldstone field which generates all
\(n\)-point functions with large-charge \(Q\) operators.  A second
goal of this paper is to use the large-charge solution to compute
Schr\"odinger-symmetry breaking corrections within the large-charge
expansion. In particular, Schr\"odinger symmetry
breaking corrections to correlation functions due to a finite
scattering length in the fundamental theory of fermions
are obtained. These corrections are found to take a
universal form in the large-charge limit, and their size is estimated
using both quantum Monte Carlo simulations, and the large-$N$
expansion, where $N$ is the number of fundamental fermion flavors~\cite{Hellerman:2023myh}.
Inclusion of effective range and other shape parameter effects in the fundamental
theory, which are essential for quantifying Schr\"odinger symmetry breaking effects
in neutron matter, require hadronic modeling, and will be left to a separate publication.

\bigskip
This paper is organized as follows. Section~\ref{sec:sonwin1} reviews
an EFT of fermions which exhibits two RG fixed points which are in
correspondence with the ideal Fermi gas and with the unitary Fermi
gas. This EFT near the unitary fixed point describes neutrons
interacting at small momentum transfers. Since the unitary Fermi gas
is superfluid, it is also described by an EFT whose degree of freedom
is the Goldstone boson of spontaneously broken particle
number. Aspects of this EFT that are relevant for what follows are
also reviewed in this section. Section~\ref{sec:corrfun} focuses on
the evaluation of non-relativistic conformal field theory correlation
functions. Schr\"odinger symmetry constraints on two- and three-point
functions are reviewed, and the technique of evaluating correlation
functions via operator insertions is introduced using the example of a
two-point correlation function of non-interacting bosons. As a
necessary prelude to what follows, the Euclidean formulation of the
superfluid EFT, which contains various subtleties, is developed. Then,
the large-charge master field solution is found and the two-, three-
and general $n$-point correlation functions with large-charge
insertions are evaluated in the large charge limit, finding results
consistent with the state-operator correspondence. In
Section~\ref{sec:sbe} Schr\"odinger symmetry breaking effects are
considered. First, these effects are formulated in the fundamental
fermion theory, and then through a simple spurion analysis are
matched to operators in the superfluid EFT. The quantitative size of these effects
is shown to follow from quantum Monte Carlo simulations.  Then, the basic structure of the
symmetry-breaking corrections is shown to be constrained by
Schr\"odinger symmetry, and, the symmetry breaking terms are derived
explicitly with the large-charge solution. Finally, in
Section~\ref{sec:large-n}, a second estimate of the size of the
symmetry-breaking corrections is provided by considering the large-$N$
expansion at large charge. Section~\ref{sec:conc} summarizes and
concludes. 

\section{Non-relativistic conformal field theories}
\label{sec:sonwin1}

\subsection{Fermions at unitarity}
\label{sec:inter}

Consider a system of spin-$1/2$ fermions which interact via two-body contact forces.
At very low energies, where derivative interactions can be ignored, the Lagrange density takes the Galilean invariant form
\begin{eqnarray}
  {\cal L}  &=&
                \psi_\sigma^\dagger \bigg\lbrack i\partial_t + \frac{\nab^{\,2}}{2M}\bigg\rbrack
                \psi_\sigma - \frac{1}{2} C_0 (\psi_\sigma^\dagger \psi_\sigma)^2 \ ,
                \label{eq:lag}
\end{eqnarray}
where the field $\psi_\sigma^\dagger$ creates a fermion of spin
$\sigma=\uparrow,\downarrow$ and $C_0$ is a bare low-energy
constant. The position-space interaction is a delta function at the origin, and therefore the
fermion interactions are highly singular at short distances.

Now consider fermion-fermion scattering. Below inelastic
thresholds, the s-wave phase shift is given by the effective range expansion
\begin{equation}
  k \cot\delta(k)\ =\  -\frac{1}{a} \ +\  \frac{1}{2} r k^2 \ +\  \mathcal{O}(k^4) \ ,
  \label{eq:eregen}
\end{equation}
where $k$ is the center-of-mass momentum, $a$ is the scattering length and $r$ is the effective range. The
EFT described by Eq.~(\ref{eq:lag}) requires
regularization and renormalization.  In dimensional regularization
with the PDS scheme~\cite{Kaplan:1998tg,Kaplan:1998we} and
renormalized at the scale $\mu$, the relation between the low-energy
constant $C_0$ and the scattering length is given by
\begin{equation}
  C_0(\mu) \ =\ \frac{4 \pi}{M} \frac{1}{{1}/{a}-{\mu}} \ .
\end{equation}
There is a fixed point at $C_0=0$, corresponding to free particles ($a=0$),
and a fixed point at $C_0=C_\star$ corresponding to a divergent scattering
length (unitarity). It is convenient to rescale the couplings to 
${\hat C}_0 \equiv C_0/C_\star$. The beta-function for the rescaled coupling
is then 
\begin{equation}
  {\hat\beta({\hat C}_0)} \ =\  \mu \frac{d}{d\mu} {\hat C}_0(\mu) \ =\ -{\hat C}_0(\mu)\left({\hat C}_0(\mu)-1\right) \ ,
  \label{eq:c0betafn}
\end{equation}
which has fixed points at ${\hat C}_0=0$ (free fermions) and $1$ (fermions at unitarity).

It is convenient to rewrite the Lagrangian with an auxiliary field $s$ as~\cite{Chowdhury:2023ahp}
\begin{eqnarray}
  \mathcal{L}  &=&
       \psi_\sigma^\dagger \bigg\lbrack i\partial_t + \frac{\nab^{\,2}}{2M}\bigg\rbrack \psi_\sigma +  \frac{1}{C_0} s^\dagger s + \psi_\downarrow^\dagger \psi_\uparrow^\dagger  s + s^\dagger \psi_\uparrow \psi_\downarrow\ .
  \label{eq:lagdimeron}
\end{eqnarray}
Then, at unitarity, the system of low-energy fermions is described by a non-relativistic conformal field theory, defined by the Lagrange density
\begin{eqnarray}
  {\cal L}_{CFT}  &=&
       \psi_\sigma^\dagger \bigg\lbrack i\partial_t + \frac{\nab^{\,2}}{2M}\bigg\rbrack \psi_\sigma +  \frac{1}{C_\star} s^\dagger s + \psi_\downarrow^\dagger \psi_\uparrow^\dagger  s + s^\dagger \psi_\uparrow \psi_\downarrow\ .
       \label{eq:lagdimeroncft}
\end{eqnarray}

Now consider a gas of fermions. At the non-interacting fixed point ($C_0=0$), the energy per particle is 
\begin{equation}
E/N \ =\  \frac{3}{5}\,\frac{\kf^2}{2M} \ ,
\label{eoNFG}
\end{equation}
where $\kf$ is the Fermi momentum. At the interacting fixed point (unitarity) there is no new scale and therefore
\begin{equation}
  E/N \ =\  \frac{3}{5}\,\frac{\kf^2}{2M}\,\xi \ ,
  \label{eq:eoNifp}
\end{equation}
where $\xi$ is a dimensionless parameter, known as the Bertsch parameter, that must be determined by numerical
simulation or via experiment (see section~\ref{sec:ufg} for details).

\subsection{Superfluid EFT at unitarity}
\label{sec:sonwin}

As the unitary limit can be approached from both negative and positive
values of $a$, it is natural to wonder: what phase does the gas of
unitary fermions find itself in the far infrared? In nuclear physics the
neutron-neutron interaction is attractive, the scattering length is
negative, and the neutron gas is superfluid. More generally, in atomic
systems the s-wave scattering length $a$ can be tuned at will via
external magnetic fields near a Feshbach resonance, and one can
smoothly crossover from a fermionic superfluid BCS state ($a < 0$) of
long-range Cooper pairs to a bosonic superfluid BEC state ($a > 0$) of
tightly-bound, repulsive dimers at the unitary fixed point. Remarkably
these seemingly very different physical systems are all described by the
same superfluid EFT in the infrared~\cite{Greiter:1989qb,Son:2005rv}.

The order parameter $\langle{\psi\psi}\rangle$ breaks the U(1)
particle-number symmetry, giving rise to a Goldstone boson excitation, $\theta(x)$,
which can be defined as (one half of) the phase of the condensate,
\begin{equation}
  \langle{\psi\psi}\rangle \ =\  |\langle{\psi\psi}\rangle|\, e^{-2i\theta} \ .
  \label{eq:GBdefined}
\end{equation}
A fundamental Galilean-invariant building block in the superfluid EFT is the field
\begin{eqnarray}
  X \ =\ D_t\theta -\ \frac{(\partial_i\theta)^2}{2M} \ ,
  \label{eq:bbgalwithtrap}
\end{eqnarray}
where $D_t\theta=\dot\theta \ -\ A_0$, with $\dot\theta=\partial_t\theta$ and $A_0$ is an external field. The effect of an external trapping potential can be included by setting
$A_0=M \omega^2 r^2/2$. It is straightforward to find the leading-order (LO) Lagrange density corresponding to fermions at unitarity~\cite{Son:2002zn,Son:2005rv},
\begin{eqnarray}
  {\cal L}_{LO} \ =\ c_0\,M^{3/2} X^{5/2}  \ ,
  \label{eq:Lfunit}
\end{eqnarray}
where $c_0$ is a low-energy constant.
Note that assigning time, $t$, scaling dimension $-2$, and space, $x_i$, scaling dimension $-1$, i.e.,
\begin{eqnarray}
\left( t\, ,\, x_i \right) \to \left( e^{2\alpha}t\, ,\, e^{\alpha}x_i \right) \ ,
  \label{eq:bsccd}
\end{eqnarray}
a scale-invariant action implies a Lagrange density with scaling dimension $5$. As $X$ has scaling dimension $2$, the Lagrange density of Eq.~(\ref{eq:Lfunit}) 
is scale invariant, and can be further shown to respect the full Schr\"odinger symmetry~\cite{Son:2005rv}.

The \ac{eom} for $\theta$ is
\begin{eqnarray}
  \partial_t \left( X^{3/2}\right) \ =\   \frac{1}{M} \partial_i \left(  \partial_i\theta X^{3/2}\right) \ .
  \label{eq:eom}
\end{eqnarray}
In spite of the apparently simple form, this is a highly non-linear equation.

The simplest non-trivial solution $\theta=\mu\, t$ gives the ground state of a system with chemical potential $\mu$.
If one expands as
\begin{equation}
  \theta(x) \ =\ \mu\, t \; -\; \phi(x) \ ,
  \label{eq:phonondefined}
\end{equation}
the fluctuations around $\phi$ correspond to phonon excitations.
The density of the system is
\begin{eqnarray}
\rho \ =\ \frac{\delta {\cal L}}{\delta X} =  \frac{5}{2}c_0 M^{3/2} X^{3/2}  \, ,
    \label{eq:dens1}
\end{eqnarray}
and one sees that the \ac{eom} takes the form of the continuity equation for the particle number symmetry:
\begin{eqnarray}
  \partial_t {\rho} \ -\ \frac{1}{M}   \partial_i\left( \partial_i \theta \rho\right)& =& 0 \ .
  \label{eq:eomLO}
\end{eqnarray}
Identifying the Lagrange density at the ground-state solution with the grand potential, $\Omega(\mu)$, and 
matching with Eq.~(\ref{eq:eoNifp}), one finds
\begin{eqnarray}
  c_0  \ =\ \frac{2^{5/2}}{15\pi^2 \xi^{3/2}} \ .
  \label{eq:c0}
\end{eqnarray}
The Hamiltonian density is related to the Lagrange density in the usual way,
\begin{eqnarray}
  {\cal H} &=& \dot\theta \frac{\partial{\cal L}}{\partial\dot\theta} \ -\ {\cal L} \ =\ \left( X -A_0- \frac{(\partial_i\theta)^2}{2M} \right)\rho-{\cal L}\ .
               \label{eq:hdensitydef}
\end{eqnarray}
The classical solution $\phi=0$, $X=\mu$ with $A_0=0$ then recovers the standard thermodynamic relation between the energy density and the grand potential.

\section{Correlation functions at large charge}
\label{sec:corrfun}

\subsection{Schr\"odinger symmetry constraints}
\label{sec:ssc}

Schr\"odinger symmetry places strong constraints on the form of the correlation functions that can be computed in the EFT~\cite{Hagen:1972pd,Niederer:1972zz,Henkel:1993sg,Mehen:1999nd,Henkel:2003pu}.
Consider the $N$-point correlation function of scalar primary operators, ${\cal O}_i$,
\begin{equation}
  G(x_1,x_2,\ldots,x_N) \ =\ -i \ev{ 0| T\left({\cal O}_1(x_1) {\cal O}_2 (x_2)\ldots{\cal O}_N(x_N) \right) |0 } \ .
  \label{eq:ssc1}
\end{equation}   
In Euclidean space, Schr\"odinger symmetry constrains the form of the two-point function as~\cite{Hagen:1972pd,Niederer:1972zz,Henkel:1993sg,Henkel:2003pu}
\begin{eqnarray}
  G(x_1,x_2) & =&  \delta_{Q_1,Q_2} \delta_{\Delta_1,\Delta_2}\, \theta(\tau_{12})\, \tau_{12}^{-\Delta_1}\,  \exp\left( -\frac{Q_1 M {\bf x}^2_{12}}{2\tau_{12}}\right) \Psi_2 + \ldots ,
                  \label{eq:ssc2}
\end{eqnarray}
where $\Psi_2$ is a constant normalization factor, and the dots corresponds to the other time ordering. 
The notation is given by $\tau_{ij}\equiv \tau_i-\tau_j$, ${\bf x}_{ij}\equiv {\bf x}_i-{\bf x}_j$.
Similarly, the three-point correlation function of scalar primary operators takes the form
\begin{eqnarray}
  G(x_1,x_2,x_3) & =& \delta_{Q_1+Q_2,-Q_3}\, \theta(\tau_{13})\, \theta(\tau_{23})\, \tau_{13}^{-\Delta_{13,2}/2}\,\tau_{23}^{-\Delta_{23,1}/2}\, \tau_{12}^{-\Delta_{12,3}/2} \nonumber \\
                 &&\times \exp\left( -\frac{Q_1 M {\bf x}^2_{13}}{2\tau_{13}}-\frac{Q_2 M {\bf x}^2_{23}}{2\tau_{23}} \right) \Psi_3(v_{123}) + \ldots ,
                    \label{eq:ssc3}
\end{eqnarray}
where $\Psi_3$ is a non-universal function of the Schr\"odinger-invariant variable $v_{123}$, which is given by
\begin{equation}
  v_{123} \equiv \frac{1}{2}\frac{\left({\bf x}_{13}\tau_{23}-{\bf x}_{23}\tau_{13}\right)^2}{\tau_{12}\tau_{13}\tau_{23}} \ =\ \frac{1}{2}\left(\frac{{\bf x}^2_{23}}{\tau_{23}}+\frac{{\bf x}^2_{12}}{\tau_{12}}
    -\frac{{\bf x}^2_{13}}{\tau_{13}} \right) \ ,
  \label{eq:ssc4}
\end{equation}   
and $\Delta_{ij,k}\equiv \Delta_i-\Delta_j-\Delta_k$.
The Ward identity constraints become less and less strong for higher
\(n\)-point correlators, which include non-universal functions of the
\((n-2)(n-1)/2\) Schrödinger-invariant variables that generalize the
\(v_{123}\)~\cite{Volovich:2009yh}.

If the correlation functions describe free-particle propagation, then
the conformal dimensions are given by their naive scaling
values. However, for interacting non-relativistic conformal field
theories, the conformal dimensions are generally strong-interaction
physics and must be obtained from either the state-operator
correspondence or an exact evaluation of the correlator.

\subsection{Free theory with operator insertions}
\label{sec:ftwoi}

A useful tool for evaluating correlation functions in the
semi-classical approximation involves exponentiating the operator
insertions and defining a new effective action, which in turn is
evaluated at its saddle point to generate the correlation
functions. This method has been used in the context of the
large-charge expansion in
Refs.~\cite{Hellerman:2017sur,Arias-Tamargo:2019xld,Orlando:2019skh,Badel:2019khk,Gaume:2020bmp}.
Here, as an example for what follows, it will be applied to the
two-point function of free, non-relativistic, boson operators. The correlation function of interest is
\begin{equation}
  G(x_1,x_2) \ =\ -i \langle 0| T\left({\cal O}_Q(x_1) {\cal O}_Q^\dagger (x_2)\right) |0\rangle \ ,
  \label{eq:ftwoi1}
\end{equation}   
where ${\cal O}_Q^\dagger$ is a primary operator of number charge $Q\equiv Q_{\cal O^\dagger}=-Q_{\cal O}\geq 0$, and scaling dimension $\Delta_Q$. 
Assuming that ${\cal O}_Q|0\rangle =0$, and $x_1=(\tau_1,{\bf x}_1)$, $x_2=(\tau_2,{\bf x}_2)$, one can write
\begin{equation}
  G(x_1,x_2) \ =\ -i \theta(\tau_{12}) \langle 0| {\cal O}_Q(\tau_1,{\bf x}_1) {\cal O}_Q^\dagger (\tau_2,{\bf x}_2) |0\rangle \ .
  \label{eq:ftwoi2}
\end{equation}   
Consider the path integral representation of the two-point function, 
\begin{equation}
  G(x_1, x_2) \ =\ \int {\cal D}\phi {\cal D}\phi^\dagger\, {\cal O}_Q(x_1) {\cal O}_Q^\dagger (x_2)\, e^{-\int \dd^4{x} {\cal L}(x)} \ ,
  \label{eq:trickfft1}
\end{equation} where now the Euclidean space formulation\footnote{The Euclidean
  space formulation is obtained via transforming $t\to -i\tau$,
  accompanied by ${\cal L}\to-{\cal L}$.} of the free theory
will be used, with a chemical potential $\mu$, 
\begin{eqnarray}
  {\cal L}  &=&
                \phi^\dagger \bigg\lbrack \partial_\tau - \frac{\nab^{\,2}}{2M} - \mu \bigg\rbrack  \phi  \ ,
                \label{eq:trickfft2}
\end{eqnarray}
where the field $\phi^\dagger$ creates a boson with unit charge. An operator of charge $Q$ can be written as
\begin{align}
  {\cal O}_Q &= {\cal N} (\phi)^{Q} \ , &  {\cal O}_Q^\dagger &= {\cal N} (\phi^\dagger)^{Q} \ ,
                                                       \label{eq:trickfft3}
\end{align}   
where ${\cal N}$ is a normalization factor.
With $\Delta_\phi=3/2$, one has $\Delta_Q=3Q/2$. Note that since ${\cal O}_Q$ is a primary operator with ${\cal O}_Q|0\rangle =0$, there is always an overall $\theta(\tau_{12})$
multiplying the two-point function which is ignored in what follows by assuming that $\tau_1>\tau_2$.

Exponentiating the insertions gives the new action,
\begin{equation}
S \ =\ \int \dd^4{x} \big\lbrack {\cal L} \ -\ {Q}\log(\phi) \delta^4(x-x_1)
\ -\ {Q} \log(\phi^\dagger) \delta^4(x-x_2) \big\rbrack \ .
\label{eq:trickfft4}
\end{equation}   
Varying this action then gives the two \ac{eom} about the free fixed point,
\begin{equation}
\phi^\dagger \bigg\lbrack \partial_\tau - \frac{\nab^{\,2}}{2M} - \mu \bigg\rbrack \phi \, =\, +{Q}\,  \delta^4(x-x_2) \ \ ,\ \
\phi \bigg\lbrack \partial_\tau + \frac{\nab^{\,2}}{2M} + \mu \bigg\rbrack \phi^\dagger \, =\, -{Q}\,  \delta^4(x-x_1)\ .
\label{eq:trickfft5}
\end{equation}   
In order to solve these equations, consider the diffusion equation
\begin{equation}
 \bigg\lbrack \partial_\tau - \frac{\nab^{\,2}}{2M} - \mu \bigg\rbrack {\cal G}(x;x_2) \, =\, \delta^4(x-x_2) \ ,
\label{eq:trickfft6}
\end{equation}   
which is satisfied by the Green's function
\begin{eqnarray}
{\cal G}(x;x_2) &=& \theta(\tau-\tau_2)\left( \frac{M}{2\pi(\tau-\tau_2)}\right)^{3/2}\exp{\left(-\frac{M({\bf x}-{\bf x}_2)^2}{2(\tau-\tau_2)}\right)} \exp{\left(\mu(\tau-\tau_2)\right)} \ ,
\label{eq:trickfft7}
\end{eqnarray}   
and, likewise,
\begin{equation}
 \bigg\lbrack \partial_\tau + \frac{\nab^{\,2}}{2M} + \mu \bigg\rbrack {\cal G}(x_2;x) \, =\, -\delta^4(x-x_2) \ .
\label{eq:trickfft7a}
\end{equation}   
The solution of the \ac{eom} in the presence of the sources is thus found to be
\begin{eqnarray}
\phi & =& i \left({Q}\right)^{1/2}\frac{{\cal G}(x;x_2)}{{\cal G}(x_1;x_2)^{1/2}} \ \ ,\ \
\phi^\dagger \ =\ -i \left({Q}\right)^{1/2}\frac{{\cal G}(x_1;x)}{{\cal G}(x_1;x_2)^{1/2}} \ .
\label{eq:trickfft8}
\end{eqnarray}   
Note that the density is given by
\begin{equation}
\rho \ =\ -\frac{\delta {\cal L}}{\delta \mu} \ =\ \phi^\dagger\phi \ =\ Q\,\frac{{\cal G}(x_1;x){\cal G}(x;x_2)}{{\cal G}(x_1;x_2)} \ ,
\label{eq:trickfft8a}
\end{equation}
which has support in the range $\tau_2\leq \tau\leq \tau_1$. Using the identity
\begin{equation}
{{\cal G}(x_1;x_2)} \ =\ \int d^3{\bf x}\,{{\cal G}(x_1;x){\cal G}(x;x_2)} \ ,
\label{eq:trickfft8b}
\end{equation}
one finds
\begin{eqnarray}
Q  &=& \int d^3{\bf x}\,\rho \, ,
\label{eq:trickfft8c}
\end{eqnarray}   
as expected.

The action at the saddle point is
\begin{equation}
S \ = \ Q \ - \ Q \log\left(-\frac{1}{2}\,Q\,{\cal G}(x_1;x_2)\right)\ ,
\label{eq:trickfft10}
\end{equation}   
and, finally,
\begin{equation}
G(x_1,x_2) \ =\  \theta(\tau_{12})\tau_{12}^{-3Q/2}\exp{\left(-\frac{M Q{\bf x}_{12}^2}{2\tau_{12}}\right)}\, \Psi_2\, e^{\mu Q\tau_{12}} \ ,
\label{eq:trickfft11}
\end{equation}   
where
\begin{equation}
\Psi_2 \ =\  {\cal N}^2 \Bigg\lbrack \left( \frac{M}{2\pi}\right)^{3/2}\,e^{1-i\pi} 2 Q\Bigg\rbrack^Q \  .
\label{eq:trickfft12}
\end{equation}   
This result is in the Schr\"odinger form of Eq.~(\ref{eq:ssc2}) only
if $\mu=0$, as the chemical potential is a Schr\"odinger-breaking
effect.

Before moving on to the more interesting interacting case, consider several
general remarks regarding this result.  
An operator of charge \(Q\) has been inserted at \(\tau = \tau_2\) and then
removed at \(\tau = \tau_1 > \tau_2\).  By construction this is the
Euclidean time-ordered correlator.  In this semiclassical calculation,
the ordering (represented here by the Heaviside function
\(\theta(\tau_{12})\)) comes from the fact that the \ac{eom} is a
diffusion equation, for which the initial-value problem is well-posed
for \(\tau > 0\) (as opposed to the backward equation for which the
problem is generically not well-posed).

\subsection{Two-point function at large charge}
\label{sec:swseft}

Before proceeding to computing the two-point function in the large-charge EFT, consider the Euclidean formulation of the superfluid EFT.
The partition function is
\begin{eqnarray}
Z \ =\  \int \DD{\theta} \exp\left(-S \right) \ =\ \int {\cal D}\theta \exp\left(-\int \dd{\tau} \dd^3{\bf x} {\cal L} \right) \ ,
\label{eq:efoeft1}
\end{eqnarray}
where, at LO, 
\begin{eqnarray}
  {\cal L}_{LO} \ =\ -c_0\,M^{3/2} X^{5/2}  \ ,
\label{eq:efoeft3}
\end{eqnarray}
with
\begin{eqnarray}
X \ =\ i \partial_\tau \theta \, -\,\ A_0\, -\,\frac{(\partial_i\theta)^2}{2M} \, .
\label{eq:efoeft4}
\end{eqnarray}
Hermiticity requires that $\theta$ be pure imaginary. Note that now
\begin{align}
	\rho & =\ -\frac{\delta {\cal L}}{\delta X} \  , & {\cal H} & =\ {\cal L} \; -\; \dot\theta \frac{\partial{\cal L}}{\partial\dot\theta} \ ,
\label{eq:efoeft7}
\end{align}  
where here $\dot\theta\equiv \partial_\tau\theta$.
The Euler--Lagrange equations take the form
\begin{eqnarray}
    \partial_\tau\left( i \frac{\delta {\cal L}}{\delta X}\right) \ =\   \frac{1}{M} \partial_i \left({\partial_i\theta}\frac{\delta {\cal L}}{\delta X}\right) \,+\, \frac{\delta{\cal L}}{\delta\theta} \ .
\label{eq:efoeft5}
\end{eqnarray}
At LO, one finds the ground-state solution
\begin{align}
	\theta  &= -i\,\mu\,\tau\  , & X &  = \ \mu\, -\,A_0 \ .
\label{eq:efoeft6}
\end{align}
For what follows, it is interesting to note that there is also a space-time dependent solution to the \ac{eom} (now taking $A_0=0$),
\begin{align}
	\theta_f(\tau,{\bf x})  &= -i\frac{M{\bf x}^2}{4 \tau}  \  , &  X_f(\tau,{\bf x}) &  = \ - \frac{M{\bf x}^2}{8\tau^2} \ .
  \label{eq:efoeft8}
\end{align}
As $X_f$ is negative definite, this solution corresponds to an unstable configuration. As will be seen below, there is a
sense in which this is the ``free'' solution, which is unsurprisingly not stable in the superfluid EFT.

\paragraph{The saddle configuration.}

Now, take $A_0=0$.
In the EFT every operator can be written as a function of the Goldstone field by simply identifying its quantum numbers.
If follows that a generic primary operator with charge $Q$ and dimension $\Delta$ must have the form
\begin{equation}
{\cal O}_{\Delta,Q} = {\cal N} \left( \frac{2}{\gamma} \right)^{\Delta/2} X^{\Delta/2} \exp\left(i Q \theta  \right)\, ,
\label{eq:swseft2}
\end{equation}   
where ${\cal N}$ is a normalization constant, and the other constant factor has been chosen for future convenience and will be defined below. Define the two-point function
\begin{equation}
G_Q(x_1,x_2) \ =\  \int \DD{\theta} {\cal O}_{\Delta,Q}(x_2) {{\cal O}}_{\Delta,-Q} (x_1)\, e^{-\int \dd^4{x} {\cal L}_I} \ ,
\label{eq:swseft3}
\end{equation}   
where ${\cal L}_I$ is the interaction Lagrange density.
In this semiclassical computation, only one time ordering will appear as a consequence of the fact that the \ac{eom} are not time-reversal invariant, as was already observed in the free case above.
In what follows, it will be assumed that  $\tau_1>\tau_2$. 

The operator insertions can be exponentiated to give
\begin{equation}
G_Q(x_1,x_2) \ =\ {\cal N}^2 \left( \frac{2}{\gamma} \right)^{\Delta} \int \DD{\theta} e^{-\int \dd^4{x} {\cal L}} \ ,
\label{eq:swseft4}
\end{equation}   
where
\begin{equation}
\mathcal{L} = \mathcal{L}_I  -  \frac{\Delta}{2}\log X \big\lbrack\delta^4(x-x_2)+\delta^4(x-x_1)\big\rbrack
   -   i Q\,\theta  \big\lbrack\delta^4(x-x_2)-\delta^4(x-x_1)\big\rbrack.
\label{eq:swseft5}
\end{equation}
Now the saddle point solution must be found in the presence of the sources. The Euler--Lagrange equation is 
\begin{eqnarray}
    \partial_\tau \rho \, +\,   \frac{1}{M} \partial_i \left(i {\partial_i\theta}\rho \right) \,=\,
Q\,\big\lbrack\delta^4(x-x_2)-\delta^4(x-x_1)\big\rbrack \,.
    \label{eq:swseft7c}
\end{eqnarray}
This is the continuity equation in the presence of a source. It may
appear confusing that there are now two distinct expressions for the
density. First, there is the density as an observable quantity related
to the specific Lagrange density which describes the system, given by
Eq.~(\ref{eq:efoeft7}), and which satisfies Eq.~(\ref{eq:swseft7c}) in
the presence and absence of sources.  On the other hand, there is the
density -- proportional to $Q$ -- which solves Eq.~(\ref{eq:swseft7c})
in the presence of sources and vanishes in the absence of
sources. This second density is a distribution and is not
directly equal to the observable density. However, the
densities integrated over space must be equal, and indeed that
equality fixes the charge $Q$ of the superfluid system.

In order to solve Eq.~(\ref{eq:swseft7c}) for $\theta=\theta(\tau,{\bf x})$, consider the ansatz
\begin{eqnarray}
i {\partial_i\theta_\epsilon} \rho_\epsilon \,=\,\epsilon \frac{1}{2} \partial_i \rho_\epsilon \ ,
    \label{eq:swseft8}
\end{eqnarray}
where $\epsilon=\pm$. This equation is readily integrated to give
\begin{eqnarray}
\rho_\epsilon \ =\  f_\epsilon(\tau)  \exp\left(2i \epsilon\theta_\epsilon\right) \ ,
    \label{eq:swseft11}
\end{eqnarray}
where $f_\epsilon(\tau)$ is an arbitrary real function of $\tau$.

\begin{figure}[!ht]
\centering
\includegraphics[width = 0.45\textwidth]{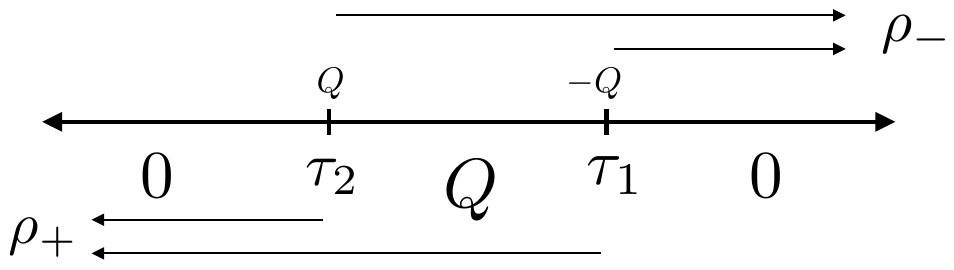}
\caption{Euclidean timeline with source charge placements (above line) and charge regions (below line).}
  \label{fig:2ptrho}
\end{figure}

Now plugging the ansatz, Eq.~(\ref{eq:swseft8}), into the \ac{eom} gives
\begin{equation}
 \bigg\lbrack \partial_\tau + \epsilon\frac{\nab^{\,2}}{2M} \bigg\rbrack \rho_\epsilon \, =\, Q\,\big\lbrack\delta^4(x-x_2)-\delta^4(x-x_1)\big\rbrack \ ,
    \label{eq:swseft13}
\end{equation}   
and it is clear that the ansatz restricts the initial highly non-linear \ac{eom} to a diffusion equation, whose solution has been shown above to be  Schr\"odinger invariant.
Noting that the diffusion equation
\begin{equation}
 \bigg\lbrack \partial_\tau - \frac{\nab^{\,2}}{2M} \bigg\rbrack \mathbb{g}(x;x') \, =\, \delta^4(x-x') 
\label{eq:swseft14}
\end{equation}   
is solved by the Green's function
\begin{eqnarray}
\mathbb{g}(x;x') &=& \theta(\tau-\tau')\left( \frac{M}{2\pi(\tau-\tau')}\right)^{3/2}\exp{\left(-\frac{M({\bf x}-{\bf x}')^2}{2(\tau-\tau')}\right)} \ ,
\label{eq:swseft15}
\end{eqnarray}   
and 
\begin{equation}
 \bigg\lbrack \partial_\tau + \frac{\nab^{\,2}}{2M} \bigg\rbrack \mathbb{g}(x';x) \, =\, -\delta^4(x-x') \ ,
\label{eq:swseft16D}
\end{equation}   
one finds the solutions
\begin{eqnarray}
\rho_- &=& +Q\big\lbrack\, \mathbb{g}(x;x_2) -\mathbb{g}(x;x_1)\,\big\rbrack \ \ , \ \
\rho_+ \;=\; -Q\big\lbrack\, \mathbb{g}(x_2;x) -\mathbb{g}(x_1;x)\,\big\rbrack  \ .
\label{eq:swseft16E}
\end{eqnarray}   
Integrating both sides over space gives
\begin{eqnarray}
\mkern-45mu\int d^3{\bf x}\,\rho_- &=& +Q\big\lbrack\, \theta(\tau-\tau_2) -\theta(\tau-\tau_1)\,\big\rbrack \ \; ,\ \;
\int d^3{\bf x}\,\rho_+ \; =\; -Q\big\lbrack\, \theta(\tau_2-\tau) -\theta(\tau_1-\tau)\,\big\rbrack .
\label{eq:swseft16A}
\end{eqnarray}   
And, as expected,
\begin{eqnarray}
Q  &=& \frac{1}{(\tau_1-\tau_2)}\int_{\tau_2+\varepsilon}^{\tau_1-\varepsilon}\int d^3{\bf x}\,\rho_\pm \ .
\label{eq:swseft16B}
\end{eqnarray}   
The temporal structure of the solution is illustrated in Fig.~\ref{fig:2ptrho}, which makes clear that the
solution has non-vanishing charge only in the interval between $\tau_2$ and $\tau_1$.

Consider the solution $\rho_-$ in the range $\tau_2+\varepsilon<\tau<\tau_1-\varepsilon$. One has
\begin{eqnarray}
\rho_- &=& Q \left( \frac{M}{2\pi(\tau-\tau_2)}\right)^{3/2}\exp{\left(-\frac{M({\bf x}-{\bf x}_2)^2}{2(\tau-\tau_2)}\right)} \ =\ f_-(\tau)  \exp\left(-2 i \theta_-\right) \ .
\label{eq:swseft16F}
\end{eqnarray}   
Now scale invariance suggests the power law form
\begin{eqnarray}
f_-(\tau)  &=& d\,\left(\frac{M}{2\pi(\tau-\tau_2)}\right)^{3/2 -\gamma}\; ,
\label{eq:swseft16G}
\end{eqnarray}
where $d$ is an irrelevant dimensionless constant that contributes to the overall normalization of the Green's function, and is here chosen for convenience to be $d=Q$,
and $\gamma$ is an anomalous dimension\footnote{$\gamma$ should not be confused with $\Delta$, the  conformal dimension that is computed below.}. Now if $\gamma=0$, then
one finds from Eq.~(\ref{eq:swseft16F}) the unstable solution $\theta_-=\theta_f$, given in Eq.~(\ref{eq:efoeft8}). It is in this sense that $\theta_f$ may be thought of as the ``free'' solution.
Therefore, one should expect superfluidity to be triggered by $\gamma\neq 0$. In this case, one finds from Eq.~(\ref{eq:swseft16F}) the Goldstone field
\begin{eqnarray}
\theta_-  &=& \frac{i}{2} \gamma \log\left(\frac{M}{2\pi(\tau-\tau_2)}\right) \;-\; i\frac{M({\bf x}-{\bf x}_2)^2}{4(\tau-\tau_2)} \ .
\label{eq:swseft16Hb}
\end{eqnarray}
Next consider the solution $\rho_+$ in the range $\tau_2+\varepsilon<\tau<\tau_1-\varepsilon$. One has
\begin{eqnarray}
\rho_+ &=& Q \left( \frac{M}{2\pi(\tau_1-\tau)}\right)^{3/2}\exp{\left(-\frac{M({\bf x}-{\bf x}_1)^2}{2(\tau_1-\tau)}\right)} \ =\ f_+(\tau)  \exp\left(2 i \theta_+\right) \ ,
\label{eq:swseft16Hc}
\end{eqnarray}   
where now
\begin{eqnarray}
f_+(\tau)  &=& d\,\left(\frac{M}{2\pi(\tau_1-\tau)}\right)^{3/2 -\gamma}\; .
\label{eq:swseft16Gb}
\end{eqnarray}
And the second solution for the Goldstone field is
\begin{eqnarray}
\theta_+  &=& -\frac{i}{2} \gamma \log\left(\frac{M}{2\pi(\tau_1-\tau)}\right) \;+\; i\frac{M({\bf x}-{\bf x}_1)^2}{4(\tau_1-\tau)} \ .
\label{eq:swseft16Hd}
\end{eqnarray}
Both solutions, Eq.~(\ref{eq:swseft16Hb}) and
Eq.~(\ref{eq:swseft16Hd}), satisfy the homogeneous \ac{eom}. However,
the inhomogeneous solution, and therefore the general solution, must
depend on both source locations. Indeed, one expects that the solution
will be invariant with respect to $Q \to - Q$ and $x_2 \leftrightarrow
x_1$. The only linear combination of $\theta_-$ and $\theta_+$ which
is invariant with respect to $x_2 \leftrightarrow x_1$ is the
sum\footnote{Invariance follows if one also takes $\gamma \to -
  \gamma$, however note that the \ac{eom} is solved
  independently of the value of $\gamma$, so the sign of this term is
  irrelevant until one fixes the charge.} $\theta_-+\theta_+$, and
indeed the sum is evidently the only linear combination that solves
the homogeneous \ac{eom}.

Finally, the general solution to the \ac{eom} and the saddle point location in the presence of sources is given by
\begin{eqnarray}
  \theta_s(\tau, {\bf x})&=&  \frac{i}{2} \gamma \log\left(\frac{\tau_1-\tau}{\tau-\tau_2}\right) \;-\; \frac{i}{4}M
  \Bigg\lbrack  \frac{({\bf x}-{\bf x}_2)^2}{(\tau-\tau_2)}\ -\ \frac{({\bf x}-{\bf x}_1)^2}{(\tau_1-\tau)}   \Bigg\rbrack \ .
\label{eq:swseft16He}
\end{eqnarray}
As will be seen below, this is the large-charge master field.  The
solution is well-defined and physical for \(\tau_2 < \tau <
\tau_{1}\).  Just as in the free case, the structure of the \ac{eom}
imposes a time ordering.  To arrive at this solution, the assumption
of Schr\"odinger invariance has entered via the diffusion-equation
ansatz of Eq.~(\ref{eq:swseft8}).  However, it is straightforward to
see that the master field solution is present in the Schr\"odinger
algebra, and this will be shown in detail below%
\footnote{This solution is general in the following sense.
  Consider a Schrödinger-invariant system and let \(\theta\) be the field that transforms with a shift under the action of the particle number operator. Then there always exists a solution to the \ac{eom} in which \(\theta\) has the form in Eq.~(\ref{eq:swseft16He}). For example, for the free theory, a special case of this solution has been found in~\cite{Son:2021kkx} as an instanton which describes the decay of a metastable bosonic droplet.}.
From the saddle solution, one finds
\begin{eqnarray}
  X[\theta_s]  &=& \frac{1}{2} \gamma \frac{(\tau_1-\tau_2)}{(\tau-\tau_2)(\tau_1-\tau)}\ -\   \frac{1}{8} M
  \Bigg\lbrack  \frac{({\bf x}-{\bf x}_2)}{(\tau-\tau_2)}\ -\ \frac{({\bf x}-{\bf x}_1)}{(\tau-\tau_1)}\Bigg\rbrack^2 \ .
\label{eq:swseft18}
\end{eqnarray}
It is convenient to shift the spatial variable to
\begin{eqnarray}
  {\bf x}  &=&    {\bf y}   \ +\ {\bf x}_2\left(\frac{\tau_1-\tau}{\tau_1-\tau_2}\right)\ +\ {\bf x}_1\left(\frac{\tau-\tau_2}{\tau_1-\tau_2}\right) \ .
\label{eq:swseft19}
\end{eqnarray}
It then follows that 
\begin{eqnarray}
\mkern-30mu  {\bf x}\,-\,{\bf x}_2  &=&    {\bf y}   \, +\, {\bf x}_{12} \left(\frac{\tau-\tau_2}{\tau_1-\tau_2}\right) \ \ \ ,\ \ \
  {\bf x}\,-\,{\bf x}_1  \;=\;    {\bf y}   \, -\, {\bf x}_{12} \left(\frac{\tau_1-\tau}{\tau_1-\tau_2}\right) ,
\label{eq:swseft19a}
\end{eqnarray}
and the spatial delta functions become
\begin{eqnarray}
\delta^4(x\;-\; x_i)&=& \delta^3({\bf y}) \delta(\tau\;-\;\tau_i) \ .
  \label{eq:swseft19b}
\end{eqnarray}
In the new coordinate,
\begin{eqnarray}
  \theta_s&=&  \frac{i}{2} \gamma \log\left(\frac{\tau_1-\tau}{\tau-\tau_2}\right) - \frac{i}{4}M
  \Bigg\lbrack  \frac{1}{\tau-\tau_2} \left({\bf y} + {\bf x}_{12} \left(\frac{\tau-\tau_2}{\tau_1-\tau_2}\right) \right)^2 \nonumber \\
  &&\qquad\qquad\qquad\qquad\qquad- \frac{1}{\tau_1-\tau} \left({\bf y}  - {\bf x}_{12} \left(\frac{\tau_1-\tau}{\tau_1-\tau_2}\right)  \right)^2
  \Bigg\rbrack 
  \label{eq:swseft19c}
\end{eqnarray}
and
\begin{eqnarray}
  X  &=& \frac{1}{2} \gamma \frac{(\tau_1-\tau_2)}{(\tau-\tau_2)(\tau_1-\tau)}\ -\   \frac{1}{8} M
  \Big\lbrack \frac{(\tau_1-\tau_2)}{(\tau-\tau_2)(\tau_1-\tau)}  \Big\rbrack^2\,{\bf y}^2 \ .
\label{eq:swseft20}
\end{eqnarray}
Note that now all dependence of the correlation function on the spatial separation is contained in the $Q$ insertion in the action.
The emergent harmonic trap is now seen by writing
\begin{eqnarray}
  X  &=& \bar\mu(\tau) \;-\; \frac{1}{2} M {\bar\omega}(\tau)^2 r^2 \; ,
\label{eq:swseft24}
\end{eqnarray}
where $r=|{\bf y}|$, and
\begin{eqnarray}
\bar\omega(\tau) &=& \frac{1}{2} \frac{(\tau_1-\tau_2)}{(\tau-\tau_2)(\tau_1-\tau)} \ \ \ , \ \ \ \bar\mu(\tau) \;=\; \frac{1}{2} \gamma \frac{(\tau_1-\tau_2)}{(\tau-\tau_2)(\tau_1-\tau)}\ .
\label{eq:swseft24B}
\end{eqnarray}
Here, it is clear that the time-dependent function $\bar\mu(\tau)$,
via the anomalous dimension $\gamma>0$, drives spontaneous symmetry
breaking\footnote{Despite the suggestive notation, this function
  should not be viewed as a time-dependent chemical potential.}.

\paragraph{The saddle from the state-operator correspondence.}

An alternative derivation of the saddle solution, this time based on
Schrödinger invariance is obtained by comparing two equivalent frames:
\begin{itemize}
\item The Galileian frame with coordinates \((\tau, \mathbf{x})\) that was used above, in which there is no background potential and the sources are inserted at finite Euclidean time, and
\item the oscillator frame, with coordinates \((\tilde \tau, \tilde{\mathbf{x}})\) in which the sources are placed at \(\tilde{\tau} = \pm \infty\) and a background field \(A_0 = M \omega^2\mathbf{y}^2/2\) is present.
\end{itemize}
These are the two sides of the non-relativistic state-operator
correspondence~\cite{Goldberger:2014hca}, in which the conformal
dimension \(\Delta\) of a primary in the Galileian frame is mapped to
the energy in units of \(\omega\), \(\tilde E/\omega\) of a state in
the oscillator frame.

The two frames are completely equivalent for Schrödinger-invariant systems and are related by
\begin{align}
  \label{eq:oscillator-to-flat}
  \begin{dcases}
    \omega \tau = \tanh(\omega \tilde{\tau})\ , \\
    \mathbf{x} = \tilde{\mathbf{x}} \frac{1}{\cosh( \omega \tilde{\tau})}\ ,
  \end{dcases} &&
                 \begin{dcases}
                   \omega \tilde{\tau} = \arctanh(\omega \tau)\ , \\
                   \tilde{\mathbf{x}} = \frac{\mathbf{x}}{\sqrt{1 - \omega^2 \tau^2}}\ .
                 \end{dcases}
\end{align}
Insertions at \(\tilde{\tau} = \pm \infty\) are then mapped to \(\tau
= \pm 1/\omega\). However, since the map is singular at these points,
such insertions in the oscillator frame are always mapped to
\(\mathbf{x} = 0\) in the Galileian frame.
Then the source points are always at  \(\mathbf{x}_1 = \mathbf{x}_2 = 0\), and therefore
\(\mathbf{x} = \mathbf{y}\). In addition, $M=1$ in this section.

Primary operators of dimension \(\Delta\) and charge \(Q\) are related by
\begin{gather}
    \Op{\Delta}{Q}(\tau,\mathbf{x}) = (\cosh(\omega \tilde{\tau}))^{\Delta} \exp\bqty*{-\frac{Q\,\omega}{2} \tanh(\omega \tilde{\tau}) \tilde{\mathbf{x}}^2} \tilde{\mathcal{O}}_{\Delta,Q}(\tilde{\tau}, \tilde{\mathbf{x}})\ ,\\
    \tilde{\mathcal{O}}_{\Delta,Q}(\tilde{\tau}, \tilde{\mathbf{x}}) = \left( 1 - \omega^2 \tau^2\right)^{\Delta/2} \exp\bqty*{\frac{Q\, \omega^2}{2} \frac{\mathbf{x}^2 \tau}{1- \omega^2\tau^2}} \Op{\Delta}{Q}(\tau, \mathbf{x})\ .
\end{gather}
The Goldstone field transforms non-linearly under Schrödinger
symmetry, so to see how the ground state solution
\(\tilde{\theta}(\tilde{\tau}, \tilde{\mathbf{x}}) = -i\,\mu\,\tilde{\tau}\) is mapped to \(\theta(\tau,\mathbf{x})\), take a
generic primary in terms of the Goldstone field and use the map above:
\begin{equation}
  \tilde{\mathcal{O}}_{\Delta,Q}(\tilde{\tau}, \tilde{\mathbf{x}})  = \tilde{X}(\tilde{\tau}, \tilde{\mathbf{x}})^{\Delta/2} e^{i Q
    \tilde{\theta}(\tilde{\tau}, \tilde{\mathbf{x}})} \mapsto  \Op{\Delta}{Q}(\tau, \mathbf{x})  = X(\tau, \mathbf{x})^{\Delta/2} e^{i Q \theta(\tau, \mathbf{x})}\ .
\end{equation}
Then one finds
\begin{align}
  \tilde{\theta} = -i\, \mu\, \tilde{\tau} & \mapsto \theta = i \frac{\mu}{2 \omega} \log\pqty*{\frac{1/\omega - \tau}{\tau + 1/\omega}} - \frac{i}{4} \left( \frac{\mathbf{x}^2}{\tau + 1/\omega} - \frac{\mathbf{x}^2}{1/\omega - \tau}\right) \ ,\\
  \tilde{X} = \mu - \frac{\omega^2}{2} \tilde{\mathbf{x}}^2 & \mapsto X = \frac{\mu}{2 \omega} \left(  \frac{1}{1/\omega - \tau} + \frac{1}{\tau + 1/\omega} \right) - \frac{1}{8} \left( \frac{\mathbf{x}}{\tau + 1/\omega} + \frac{\mathbf{x}}{1/\omega - \tau}\right)^2\ ,
\end{align}
which is the field configuration generated by two insertions at \((\tau = \pm 1/\omega, {\bf x}=0)\).
These expressions are consistent with those found working directly in the Galileian frame with the identification
\begin{equation}
  \gamma = \frac{\mu}{\omega} = \frac{\mu}{2}\tau_{12}\; .
\end{equation}

\paragraph{Evaluation of the 2-point function.}

The density is given by
\begin{eqnarray}
\rho &=& -\frac{\delta {\cal L}}{\delta X} \ =\ -\frac{\delta {\cal L_I}}{\delta X} \;+\; \frac{\Delta}{2} X^{-1} \big\lbrack\delta^4(x-x_2)+\delta^4(x-x_1)\big\rbrack\ .
    \label{eq:swseft21}
\end{eqnarray}
Now, using Eq.~(\ref{eq:swseft19b}) and Eq.~(\ref{eq:swseft20}), it is easy to see that the second term has no support and therefore gives a vanishing contribution
to the density. Therefore, taking
\begin{eqnarray}
{\cal L}_I&=&  {\cal L}_{LO} \ =\ -c_0\,M^{3/2} X^{5/2}  \ \ \ , \ \ \  \rho \ =\ \frac{5}{2} c_0\,M^{3/2} X^{3/2}\ ,
    \label{eq:swseft22}
\end{eqnarray}
it is straightforward to check that $\rho$  satisfies the homogeneous \ac{eom} at the saddle-point solution, Eq.~(\ref{eq:swseft16He}).

\begin{figure}[!ht]
  \centering
  \begin{minipage}{0.45\linewidth}
    \includegraphics[width = 1\textwidth]{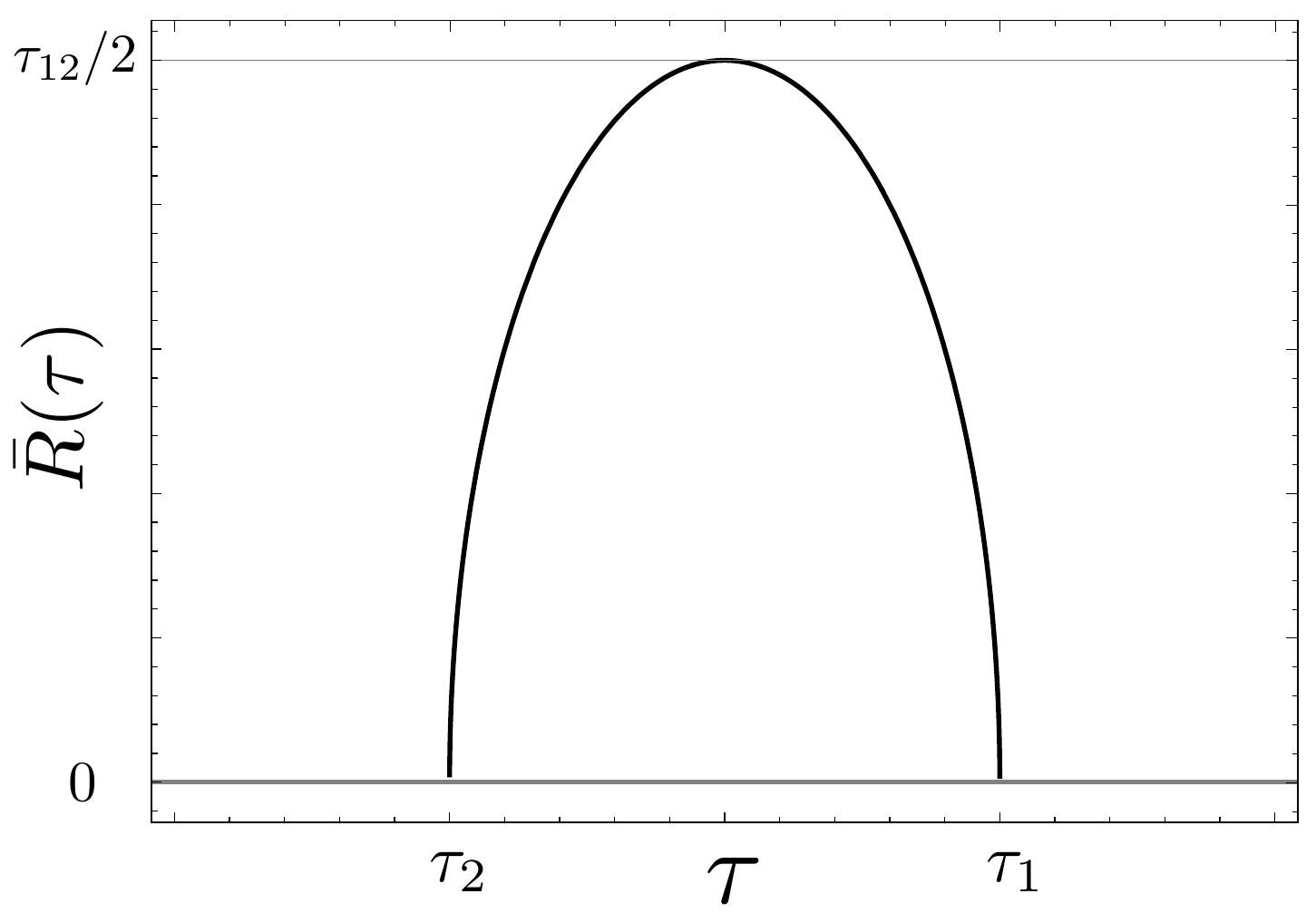}    
  \end{minipage}
  \hfill
  \begin{minipage}{0.45\linewidth}
    \includegraphics[width = 1\textwidth]{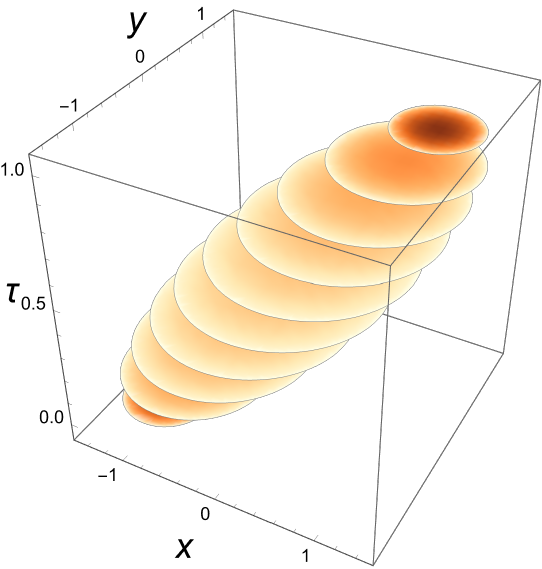}    
  \end{minipage}
\caption{Left: plot of the radius of the superfluid droplet vs Euclidean time as described in the text. Right: equal-time slices representing the evolution of the droplet size for insertions at \((0,-1,-1)\) and \((1,1,1)\). Darker colors correspond to denser regions.}
  \label{fig:Rbarvstau}
\end{figure}

To satisfy the \ac{eom} in the presence of sources, the density at the saddle-point solution must be integrated over all space to fix the charge.
Now note that the density vanishes when $r\to \bar{R}(\tau)$ with
\begin{eqnarray}
\bar{R}(\tau)& =& \sqrt{\frac{2\bar\mu}{M}}\frac{1}{\bar\omega} \ =\ \frac{2}{\sqrt{M}}\left(\gamma \frac{(\tau-\tau_2)(\tau_1-\tau)}{(\tau_1-\tau_2)}\right)^{1/2}.
\label{eq:swseft25}
\end{eqnarray}
Hence, in the presence of the sources the superfluid system behaves like a droplet whose size is governed by the Euclidean-time separation between sources.
Fig.~\ref{fig:Rbarvstau} illustrates the droplet radius forming at the source $\tau_2$, rising to its maximum at radius $\tau_{12}/2$, and returning to zero
at the sink, $\tau_1$.

One now easily finds
\begin{eqnarray}
\int d^3{\bf x}\,\rho  &=& 4\pi \int_0^{\bar R(\tau)} dr r^2 \rho \ =\ \frac{\bar\mu^3}{3\bar\omega^3 \xi^{3/2}} \ = \ \frac{1}{3\xi^{3/2}}\gamma^3 \ = \ Q \ .
\label{eq:swseft26}
\end{eqnarray}   
Therefore, the anomalous dimension is related to the charge by
\begin{eqnarray}
\gamma &=& {3^{1/3}\xi^{1/2}}Q^{1/3} \,,
\label{eq:swseft27}
\end{eqnarray}   
and 
\begin{eqnarray}
\bar\mu(\tau) \;=\; {3^{1/3}\xi^{1/2}}Q^{1/3}\bar\omega(\tau) \,.
\label{eq:swseft27B}
\end{eqnarray}
Having found the solution to the \ac{eom}, the two-point function at the saddle is straightforwardly evaluated:
\begin{equation}
    G_Q(x_1,x_2) = \eval*{ \Op{\Delta}{Q}(x_2) \Op{\Delta}{-Q} (x_1)  e^{-S[\theta]} }_{\theta = \theta_{S}} + \ldots \ ,
\end{equation}
where the dots are subleading effects in the large-charge EFT (see below).
Because of the insertions at \(\tau = \tau_2\) and \(\tau = \tau_1\), it is convenient to view this expression as
\begin{equation}
    G_Q(x_1,x_2) = \lim_{\varepsilon \to 0} \eval*{\Op{\Delta}{Q}(\tau_2 + \varepsilon, \mathbf{x}_2) \Op{\Delta}{-Q} (\tau_1 - \varepsilon, \mathbf{x}_1)  \exp\left[-\int_{\tau_2 + \varepsilon}^{\tau_1-\varepsilon} \dd{\tau} \int \dd^3{\mathbf{x}} {\cal L}_I[\theta]\right] }_{\theta = \theta_{S}} .
\end{equation}
The existence of this limit will then give a constraint that will be used to compute the conformal dimension \(\Delta\).
  
The only thing that remains to compute is the value of the action at the saddle:
\begin{equation}
  \label{eq:saddle-action}
  \begin{aligned}
    S_I &= \int_{\tau_2 + \varepsilon}^{\tau_1-\varepsilon} \dd{\tau} \int \dd^3{\mathbf{x}} {\cal L}_I = - 4 \pi c_0 \int_{\tau_2 + \varepsilon}^{\tau_1-\varepsilon} \dd{\tau} \int_0^{\bar R(\tau)} r^2 \dd{r} X(\tau, r)^{5/2}  \\
        &=  - \frac{5 c_0 \pi^2 \gamma^4}{16 \sqrt{2}}  \log\left( \frac{\tau_{12}}{\varepsilon} \right) + \mathcal{O}(\varepsilon)= - \frac{3^{1/3} \xi^{1/2}}{4} Q^{4/3} \log\left( \frac{\tau_{12}}{\varepsilon} \right) + \mathcal{O}(\varepsilon) .
  \end{aligned}
\end{equation}
Simply substituting the saddle solution into the expression for the operators one then finds
\begin{equation}
  G_Q(x_1,x_2) = \mathcal{N}^2 \varepsilon^{Q \gamma - \Delta - \frac{3^{1/3} \xi^{1/2}}{4} Q^{4/3}} \Bigg\lbrack {\tau_{12}^{-Q \gamma  +\frac{3^{1/3} \xi^{1/2}}{4} Q^{4/3} }} \exp\pqty*{-\frac{Q M \mathbf{x}_{12}^2}{2 \tau_{12}}}
  + \mathcal{O}(\varepsilon) \Bigg\rbrack\; .
\end{equation}
Cancellation of the divergence for \(\varepsilon \to 0\) requires
\begin{equation}
    \label{eq:2pt-no-divergence}
        Q \gamma - \Delta - \frac{3^{1/3} \xi^{1/2}}{4} Q^{4/3} = 0
\end{equation}
and leads to the solution
\begin{equation}
    G_Q(x_1,x_2) =  \mathcal{N}^{2} {\tau_{12}^{-\Delta}} \exp\pqty*{-\frac{Q M \mathbf{x}_{12}^2}{2 \tau_{12}}} \ ,
\end{equation}
where
\begin{equation}
    \Delta = Q \gamma - \frac{3^{1/3} \xi^{1/2}}{4} Q^{4/3} = \frac{3^{4/3}}{4} \xi^{1/2} Q^{4/3}
\end{equation}
in agreement with the state-operator correspondence large-charge result~\cite{Kravec:2018qnu}.

Note also that \(\gamma\) measures how much the conformal dimension changes when the charge changes by one unit:
\begin{equation}
    \label{eq:gamma-as-variation}
    \frac{d \Delta}{d Q} = \gamma\ .
\end{equation}
This shows that equation~\eqref{eq:2pt-no-divergence} is simply a Legendre transform relating \(\Delta \) to the value of the action at the saddle.
In the oscillator frame, this is precisely the thermodynamic relation between the free energy \(F(Q)\) and the grand potential \(\Omega(\mu)\).

\paragraph{EFT power counting.}

Note that up to this point, there has been no explicit use of the
large-charge expansion.  However, the semi-classical expansion is
used implicitly in a claim that the effective action, evaluated at the saddle, 
provides the large-charge solution. This is,
of course, true when $Q$ is large, as expected. However, it is worth
considering the various scales in the problem in some detail.
This parallels consideration of the large-charge EFT in the presence of a
harmonic trap~\cite{Hellerman:2021qzz}. However, in the state-operator
correspondence, the trap frequency, $\omega$, is an artificial
parameter which is taken to zero as $Q\to\infty$ in a manner that
leaves the density fixed. In the exact solution given here, the
emergent trap frequency, $\bar\omega$, vanishes as the temporal
separation of the sources increases. Say $\tau_1=\Delta\tau$ and
$\tau_2=-\Delta\tau$.  Then, $\Delta\tau=(\tau_1-\tau_2)/2$, and for
fixed $\tau$ and $\Delta\tau\to\infty$, one has
\begin{equation}
\bar\omega  \ \to \ \Delta\tau^{-1} \to 0 \ .
\label{eq:hofs1}
\end{equation}   
One also has
\begin{equation}
{\bar R}\ \sim\ \frac{Q^{1/6}}{M^{1/2 }\bar\omega^{1/2}}\ \ , \ \  V\ \sim\ {\bar R}^3\ \sim \ \frac{Q^{1/2}}{M^{3/2}{\bar\omega}^{3/2}}\ \ , \ \ \rho \ \sim\ \frac{Q}{V} \ \sim\ M^{3/2}\bar\omega^{3/2} Q^{1/2}\ ,
\label{eq:hofs2}
\end{equation} where $V$ is a measure of the spatial volume occupied by the
superfluid droplet.  Therefore, one can take $Q\to\infty$,
$\bar\omega\to 0$ while keeping $\rho$ fixed if in the power counting
$Q\sim M^3 \Delta\tau^3$.  Then, taking the \ac{ir} scale to be $p_{IR}\equiv{\bar R}^{-1}$
and the \ac{uv} scale to be $\Lambda_{UV}\equiv\rho^{1/3}$, the large-charge EFT can be organized as an
expansion in the ratio $p_{IR}/\Lambda_{UV} \sim Q^{-1/3}$ which
is small when $Q$ is large. This power counting argument provides a
basis for organizing operators and higher-order corrections in the large-charge EFT.

\paragraph{NLO corrections.}

The easiest way to identify the NLO corrections for the two-point function result is to write the next terms in the EFT in the oscillator frame.
The reason is that there is a natural \ac{ir} scale \(p_{IR}^2 \sim \omega\) given by the oscillator frequency and a natural \ac{uv} scale \(\Lambda_{UV}^2 \sim \mu\).
The leading scale-invariant terms, up to field redefinitions, can be chosen to be (with $M=1$)
\begin{equation}
  \label{eq:Lnlofunit}
  \mathcal{L}_{osc} = c_0 X^{5/2} + c_1 X^{-1/2} (\partial_i X)^2 + X^{1/2} \pqty*{ c_2 \pqty{\partial_i \partial_i \theta}^2 + c_3 (\partial_i \partial_j \theta)^2 + c_4 \partial_i \partial_i A_0}
\end{equation}
with \(A_0 = \omega^2 \mathbf{x}^2/2\).
The $c_{i}$ are low-energy constants that are determined by the
short-distance physics. On the basis of power counting,
each derivative counts as $Q^{-1/3}$, and therefore one expects NLO
contributions to the scaling dimensions suppressed by $Q^{-2/3}$ as
compared to the LO. This is indeed correct~\cite{Kravec:2018qnu}.

Fixing the charge breaks the Schrödinger symmetry and in general the EFT will not be invariant order-by-order in \(\omega/\mu\).
When passing to the flat frame, the \(c_0\) and \(c_1\) terms are separately invariant under the transformations in Eq.~\eqref{eq:oscillator-to-flat} and enter the flat-frame EFT as they are.
As for the other terms, one can verify that the computation of the conformal dimension in the oscillator frame and of the two-point function in the flat frame agree if the following condition is met:
\begin{equation}
  c_4 = - 3 c_2 - c_3 \ .  
\end{equation}
This relation agrees with the condition of conformal invariance when the LO \ac{eom} is used, as found in Ref.~\cite{Ma_es_2009}.
This is also consistent with Ref.~\cite{Hellerman:2023myh}, which considers the
superfluid EFT at large-charge and large-$N$ and finds that the explicit trap-dependent contribution to the $c_4$ operator in Eq.~(\ref{eq:Lnlofunit})
is present in the Schr\"odinger-invariant EFT. 

As is generally the case, there is large freedom in defining the fields and transformation properties in the EFT.
In the literature different choices have appeared~\cite{Son:2005rv,Ma_es_2009,Kravec:2018qnu}, but they must all ultimately agree on the resulting values for the physical observables.

\subsection{Three-point function and higher at large charge}
\label{sec:tpfeft}

Again consider the primary operator with charge $Q$ and dimension $\Delta$ given in Eq.~(\ref{eq:swseft2}), and 
now define the three-point function
\begin{equation}
G_Q(x_1,x_2,x_3) \ =\  \int \DD{\theta}  {\cal O}_{\Delta_3,Q_3}(x_3) {{\cal O}}_{\Delta_2,Q_2} (x_2) {{\cal O}}_{\Delta_1,Q_1} (x_1)\, e^{-\int \dd^4{x} {\cal L}_I} \ ,
\label{sec:tpfLO2}
\end{equation}   
where ${\cal L}_I$ is the interaction Lagrange density.
The operator charges and dimensions are chosen as follows:
  \begin{align}
    Q_3 &= Q \ , & Q_2 &= q \ , & Q_1 &= -Q - q\ , \\
    \Delta_3 &= \Delta_Q \ , & \Delta_2 &= \Delta_{q} \ , & \Delta_1 &= \Delta_{Q + q} \ ,
  \end{align}
and the large-charge approximation is taken in which \(q \ll Q\). In this limit, the
small-charge insertion is seen as a probe in the field configuration
generated by the two large charge insertions~\cite{Monin:2016jmo,Cuomo:2020rgt,Dondi:2022wli}.
Semiclassically, this requires the insertions to be ordered as \(\tau_3 < \tau_2 < \tau_1\).
Concretely, one can use the same solution that was found above,
corresponding to an insertion of charge \(Q\) %
and evaluate the three-point function at the saddle,%
\footnote{One could equivalently use the solution corresponding to charge \(Q + q\) as a background.}
\begin{figure}[!ht]
\centering
\includegraphics[width = 0.65\textwidth]{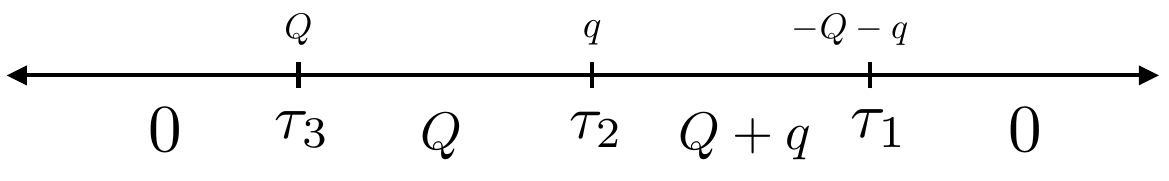}
\caption{Euclidean timeline with source charge placements (above line) and charge regions (below line).}
  \label{fig:3ptrho}
\end{figure}
\begin{multline}
      G_{Q}(x_1,x_2,x_3) = \ev{\Op{\Delta_Q}{Q}(x_3) \Op{\delta}{q}(x_2) \Op{\Delta_{Q+q}}{-Q-q}(x_1)} \\
      = \lim_{\varepsilon \to 0} \eval*{\Op{\Delta_Q}{Q} (\tau_3 + \varepsilon, \mathbf{x}_3) \Op{\delta}{q} (\tau_2, \mathbf{x}_2) \Op{\Delta_{Q+q}}{-Q-q} (\tau_1 - \varepsilon, \mathbf{x}_1) e^{-S_I[\theta]}}_{\theta=\theta_{S}} + \dots,
  \end{multline}
where, for simplicity, $\delta\equiv \Delta_q$.
The temporal structure of the solution is illustrated in Fig.~\ref{fig:3ptrho}, which makes clear that
in the large-charge limit, \(q \ll Q\), there is charge $Q$ in the interval between $\tau_3$ and $\tau_1$,
and vanishing charge elsewhere.
Note that at \(\tau = \tau_2\) the solution is regular and there is no need to move the insertion point.
The value of \(S_{I}\) is the same as was found above (with $\tau_{12}\to \tau_{13}$),
\begin{equation}
    S_I = - \frac{3^{1/3} }{4}\xi^{1/2} Q^{4/3} \log\left( \frac{\tau_{13}}{\varepsilon} \right) + \mathcal{O}(\varepsilon)  \ ,
\end{equation}
and all that remains to do is to insert the values at the saddle, observing that the master field in \(x = x_2\) can be written in terms of the Schrödinger-invariant variable \(v_{123}\):
\begin{align}
    \eval*{e^{i q \theta(x_2)}}_{\theta = \theta_s} &= \pqty*{\frac{\tau_{23}}{\tau_{12}}}^{q \gamma/2} \exp\bqty*{-\frac{1}{2} q M v_{123} + \frac{1}{2} q M \frac{\mathbf{x}_{23}^2}{\tau_{23}} - \frac{1}{4} q M \frac{\mathbf{x}_{13}^2}{\tau_{13}}}  \ , \\
  \eval*{X(x_2)}_{\theta=\theta_s} &= \frac{\gamma}{2}  \pqty*{\frac{1}{\tau_{12}} + \frac{1}{\tau_{23}}} \pqty*{1 - \frac{1}{2\gamma} M v_{123}}  \ .
\end{align}
Then one has
\begin{multline}
    G_{Q}(x_1,x_2,x_3)  = \mathcal{N}^3 \lim_{\varepsilon \to 0} \varepsilon^{- \frac{3^{1/3} \xi^{1/2}}{4} Q^{4/3} + q \gamma /2  + Q \gamma - (\Delta_{Q+q} + \Delta_Q)/2 } \\
    \times \tau_{12}^{-(q \gamma + \delta)/2} \tau_{23}^{(q \gamma - \delta)/2} \tau_{13}^{\delta/2 - 1/2(q+2 Q) + 3^{1/3}/4 \xi^{1/2} Q^{4/3}}  \\
     \times \exp\left[Q_1 M\mathbf{x}_{13}^2/(2 \tau_{13}) + Q_2 M\mathbf{x}_{23}^2/(2 \tau_{23}) \right] \\
     \times \left(1 - \frac{1}{2 \gamma} M v_{123} \right)^{\delta/2} e^{-q M v_{123}/2}\ .
\end{multline}
In this case, the cancellation of the divergence reads
\begin{equation}
      - \frac{3^{1/3} \xi^{1/2}}{4} Q^{4/3} + q \frac{\gamma}{2} + Q \gamma - \frac{\Delta_{Q+q} + \Delta_Q}{2} = 0.
\end{equation}
In the discussion of the two-point function, it was already found that \(\gamma\) measures the variation of \(\Delta\) corresponding to a (small) variation of the charge (Eq.~\eqref{eq:gamma-as-variation}),
\begin{equation}
      \Delta_{Q +q} - \Delta_Q = \gamma q \ .
\end{equation}
Putting it all together, the three-point function takes the expected form of Eq.~(\ref{eq:ssc3}), fixed by Schrödinger invariance, with
\begin{equation}
       \Psi_3(v_{123}) = \left(1 - M \frac{v_{123}}{2 \times 3^{1/3} \xi^{1/2} Q^{1/3}} \right)^{\delta/2} e^{-q M v_{123}/2}\ .
\end{equation}
This result is consistent with what was found in Ref.~\cite{Kravec:2018qnu}.

\begin{figure}[!ht]
\centering
\includegraphics[width = 0.75\textwidth]{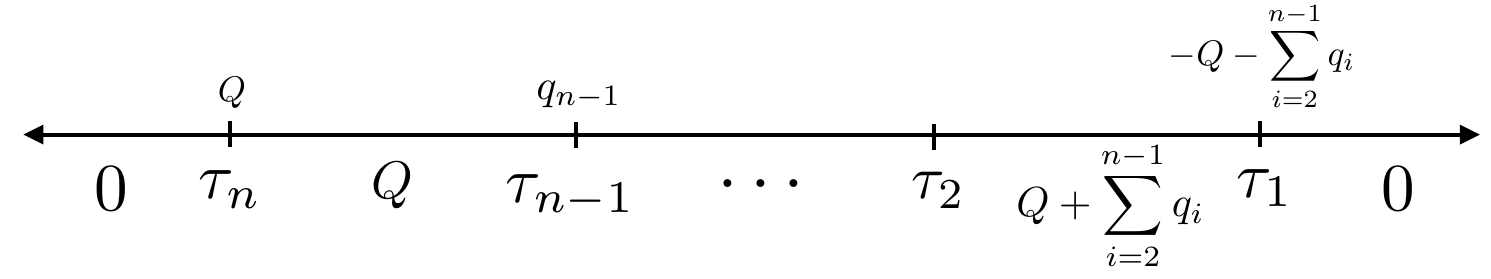}
\caption{Euclidean timeline for $n$-point function with source charge placements (above line) and charge regions (below line).}
  \label{fig:nptrho}
\end{figure}

The computation of the three-point function can be generalized in a straightforward way to any \(n\)-point function with two heavy insertions and \(n-2\) light ones:
\begin{equation}
      G_Q(x_1, \dots ,x_n) = \ev{\Op{\Delta_Q}{Q}(x_n) \Op{\delta_{n-1}}{q_{n-1}}(x_{n-1}) \dots \Op{\delta_2}{q_2}(x_2) \Op{\Delta_{Q'}}{-Q'}(x_1)} \ ,
\end{equation}
where, by charge conservation
\begin{equation}
  Q' = Q + \sum_{i=2}^{n-1} q_i = Q + q .
\end{equation}
The temporal structure of the solution is illustrated in Fig.~\ref{fig:nptrho}.
The goal is to compute this correlator semiclassically in the EFT,
treating the \(n-2\) insertions at \(x_2, \dots, x_{n-2}\) as probes
for the field profile generated by the heavy insertions at \(x_1\) and
\(x_n\) (which implies that the small insertions must occur
between \(\tau_1\) and \(\tau_n\)).  Two conditions have to be met:
\begin{itemize}
\item the inner insertions have to be light, \(|q_i| \ll Q\), and $|q| \ll Q$,
\item the distance between any two insertions has to be much larger than the \ac{uv} scale
  \begin{equation}
    \frac{1}{\tau_{ij}} \ll \Lambda_{UV} = \rho^{1/3} \approx \frac{Q^{1/3}}{\tau_{1n}} \ .
  \end{equation}
\end{itemize}
If the insertion of charge \(Q\) at \(x_n\) is taken as a reference, one recovers
precisely the same profile discussed in the previous sections and
the \(n\)-point function is simply the product of \(n\) contributions
evaluated at the saddle:
\begin{multline}
  G_Q= \lim_{\varepsilon \to 0} \Op{\Delta_Q}{Q}(\tau_n - \varepsilon, \mathbf{x}_n) \Op{\delta_{n-1}}{q_{n-1}}(x_{n-1}) \dots \\
    \Op{\delta_2}{q_2}(x_2) \Op{\Delta_{Q'}}{-Q'}(\tau_1 - \varepsilon, \mathbf{x}_1) \eval*{e^{-S_I[\theta]}}_{\theta=\theta_{S}} + \dots%
\end{multline}

The regulator \(\varepsilon\) is only needed at \(x_1\) and \(x_n\), so that the \(\varepsilon\)-dependence inside the limit is given by
\begin{equation}
  \varepsilon^{- \frac{3^{1/3} \xi^{1/2}}{4} Q^{4/3} + \gamma (Q' + Q)/2 - (\Delta_{Q'} + \Delta_Q)/2 } \ ,
\end{equation}
where the fact that the value of the action at the saddle is still given by Eq.~\eqref{eq:saddle-action} has been used.
Then the divergence is canceled if
\begin{equation}
  - \frac{3^{1/3} \xi^{1/2}}{4} Q^{4/3} + \frac{\gamma}{2}\pqty*{Q' + Q} - \frac{1}{2} \pqty*{\Delta_{Q'} + \Delta_Q} = 0\ ,
\end{equation}
which is identically true since, as seen in the computation of the two-point function,
\begin{align}
  \Delta_Q &= Q \gamma - \frac{3^{1/3} \xi^{1/2}}{4} Q^{4/3} && \text{and} && \Delta_{Q'} - \Delta_Q = \gamma \pqty*{Q'-Q}.
\end{align}
Consistently with the appropriate Ward identities, the result can be parameterized in terms of \((n-1)(n-2)/2\) independent conformally-invariant variables
\begin{align}
  v_{ijn} &\equiv  \frac{1}{2}\left(\frac{{\bf x}^2_{jn}}{\tau_{jn}} + \frac{{\bf x}^2_{ij}}{\tau_{ij}} 
-\frac{{\bf x}^2_{in}}{\tau_{in}} \right) & i < j < n\ .
\end{align}
Now one can write the general expression of the master field in terms of \(v_{1in}\):
\begin{align}
  \eval*{e^{i q_i \theta(x_i)}}_{\theta = \theta_s} &= \pqty*{\frac{\tau_{in}}{\tau_{1i}}}^{q_i \gamma/2} \exp\bqty*{-\frac{1}{2} q_i M v_{1in} + \frac{1}{2} q_i M \frac{\mathbf{x}_{in}^2}{\tau_{in}} - \frac{1}{4} q_i M \frac{\mathbf{x}_{1n}^2}{\tau_{1n}}}  \ , \\
  \eval*{X(x_i)}_{\theta=\theta_s} &= \frac{\gamma}{2} \pqty*{\frac{1}{\tau_{1i}} + \frac{1}{\tau_{in}}}\pqty*{1 - \frac{1}{2\gamma} M v_{1in}}  \ ,
\end{align}
to find that the \(n\)-point function takes a factorized form:
\begin{multline}
  G_Q = \mathcal{N}^2 \tau_{1n}^{- \Delta_Q - q \gamma/2 + \delta/2} \exp\bqty*{-(Q + q)  M \frac{\mathbf{x}_{1n}^2}{2 \tau_{1n}} } \prod_{i=2}^{n-1} \mathcal{N}_i \tau_{1i}^{-q_i \gamma/2 - \delta_i/2} \tau_{in}^{q \gamma_i/2 - \delta_i/2} \exp\bqty*{q_i M \frac{\mathbf{x}_{in}^2}{2 \tau_{in}} } \\ \left(1 - M \frac{v_{1in}}{2 \times 3^{1/3} \xi^{1/2} Q^{1/3}} \right)^{\delta_i/2} e^{-q_i M v_{1in}/2} \ ,
\end{multline}
where \(q = q_2 + \dots + q_{n-2}\) and \(\delta = \delta_2 + \dots + \delta_{n-2}\).

\section{Symmetry breaking effects}
  \label{sec:sbe}

\subsection{Fermions near unitarity}
\label{sec:ufg}

Consider now the inclusion of small Schr\"odinger-breaking effects in the fundamental theory.  Using Eq.~(\ref{eq:lagdimeron}) one can write
\begin{eqnarray}
  {\cal L}  &=&   {\cal L}_{CFT} \ +\ \frac{M}{4\pi a} s^\dagger s \ -\ \frac{M^2 r}{16\pi} s^\dagger\left( i\overleftrightarrow{\partial_t} + \frac{\nab^{\,2}+\nabl^{\,2}}{4M} \right) s \ ,
  \label{eq:lagdimerondecomp}
\end{eqnarray}
where, in addition, effective range corrections have been included~\cite{Beane:2000fi,Chowdhury:2023ahp}. The scattering length corrections therefore enter via a relevant dimension-4 operator
(the field $s$ at unitarity has dimension 2), and the effective range corrections enter via an irrelevant dimension-6 operator.

The energy per particle $E/N$ of the interacting Fermi gas in the
near-Schr\"odinger limit can be written at very-low
densities\footnote{Note that the range of validity of the EFT of
  neutron contact operators is, strictly speaking, not set by the pion
  mass, $M_\pi$, but rather by $M_\pi/2$, which is the position of the
  t-channel branch point in the neutron-neutron scattering amplitude.}
as~\cite{Bulgac:2005zza,vanKolck:2017jon}
\begin{equation}
E/N \ =\  \frac{3}{5}\, \frac{\kf^2}{2M}\,  \left(\xi \ -\  \frac{\zeta}{\kf a} \ -\  \frac{\zeta_2}{\kf^2 a^2}\ +\ \ldots \ +\ \eta\, \kf r \ + \ \ldots    \right) \ .
\label{eq:eoNunitarity}
       \end{equation}
Here the various dimensionless universal parameters have been
determined using quantum Monte Carlo simulations. From
Ref.~\cite{PhysRevA.84.061602}, $\xi = 0.372(5)$ (Bertsch parameter)
and $\eta=0.12(3)$. From the simulation data in Ref.~\cite{Chang:2004zza}, it is
straightforward to extract $\zeta =0.8(3)$ and $\zeta_2=1.0(6)$.

Note that the effective range contributions to the energy density grow
with Fermi momentum with respect to the energy at unitarity. This
implies that in the large-charge limit, the superfluid EFT is
sensitive to arbitrarily short distance scales in the fundamental
fermion theory. Hence, the effect of the effective range and all
higher shape parameter corrections that arise from operators with
arbitrary numbers of derivative interactions are of the same size,
leading to a breakdown of the fundamental-fermion EFT expansion. This
renders it difficult to assess the effect of Schr\"odinger-breaking
effects due to finite effective-range corrections in the large-charge
EFT, as evidently knowledge of the equation-of-state at high densities
is required\footnote{Recent state-of-the-art simulations of the energy
  density in neutron matter~\cite{Lim:2023dbk} indicate a remarkable
  closeness to the unitary curve up to unexpectedly high
  densities.}. This will be treated in a separate publication, and
therefore effective range and higher shape parameter corrections in
the fundamental theory are assumed to vanish in what follows.

\subsection{Symmetry breaking in EFT}
\label{sec:schrosb}

The leading conformal-breaking effects are straightforward
to include using spurion formalism.  In the fundamental theory,
recalling that the field $s$ has scaling dimension $2$, the
conformal-breaking effects can be made scale invariant by assigning
$1/a$ scaling dimension 1. Correspondingly, in the EFT, the leading conformal-breaking
effects due to a finite scattering length are encoded as\footnote{Note that this is the Euclidean Lagrange density, and therefore the
relationship with the energy density is given by Eq.~(\ref{eq:efoeft7}).}
\begin{eqnarray}
{\cal L}_{\textsc{sb}} \ =\  g_1\,a^{-1}{M X^2} \ +\ g_2\,a^{-2}M^{1/2} X^{3/2} \ ,
  \label{eq:Lsbfunit}
\end{eqnarray}
where the $g_i$ are dimensionless constants\footnote{Note that the contribution of the leading scattering length correction to the phonon dispersion relation has been considered in Ref.~\cite{Escobedo:2009bh}.}.
Matching to Eq.~(\ref{eq:eoNunitarity}) one finds
\begin{align}
g_1  \ =\   -\frac{2 \zeta}{5\pi^2 \xi^{2}}=-0.23(10)\ \ & ,&  \ \  g_2  = -\frac{\sqrt{2}}{25\pi^2 \xi^{5/2}}\left(4\zeta^2+5\zeta_2 \xi \right)= -0.34(17) \ ,
	\label{eq:g0g1}
\end{align}  
where the Monte Carlo results quoted above have been used in the numerical determination.

Note that following this procedure for the effective range
corrections, one would assign $r$ scaling dimension
$-1$. Correspondingly, the range corrections would enter perturbative
matching via an $X^3$ operator which scales as $\mu^3$.  As the LO
Schr\"odinger-invariant operator scales as $\mu^{5/2}$, formally there
can be no such operator in the large-charge limit. This suggests that
the utility of the large-charge EFT in the presence of significant
effective range, and higher shape parameter effects, will be system
specific.

\subsection{Symmetry breaking at large charge}
\label{sec:swscbe}

Remarkably, Schr\"odinger symmetry places strong constraints on the general form of the symmetry-breaking corrections
due to scattering length and effective range correction~\cite{Chowdhury:2023ahp}. In the fundamental fermion theory with
a two-point function constructed from an operator of arbitrary charge $Q$, it can be shown in conformal perturbation theory
that
\begin{eqnarray}
G_Q(x_1,x_2) &=& G^{CFT}_Q(x_1,x_2)\,\Big\lbrack1 + c' a^{-1}\,\tau_{12}^{1/2} \Big\rbrack \,,
\label{eq:swscbe1a}
\end{eqnarray}  
where $c'$ is an arbitrary constant that is not fixed by
symmetry. In conformal perturbation theory this symmetry-breaking
effect is computed by obtaining the symmetry-breaking action at the
Schr\"odinger-symmetric master field solution.

At leading order in the symmetry breaking, the saddle solution remains
unchanged.  This is reflected in the fact that in computing $n$-point
functions in the theory with symmetry breaking, in the limit in which
two insertions are much heavier than the others and the
symmetry-breaking is small, the only new effect is a variation in the
value of the action at the saddle:
\begin{equation}
  \begin{aligned}
    \ev*{\Op{Q}{\Delta} \prod_{i=2}^{n-2} \Op{\delta_i}{q_i} \Op{\Delta'}{-Q'}}_{\textsc{sb}} &\sim \eval*{\Op{Q}{\Delta} \prod_{i=2}^{n-2} \Op{\delta_i}{q_i} \Op{\Delta'}{-Q'} e^{-S[\theta] - S_{\textsc{sb}}[\theta]}}_{\theta = \theta_{S}} \\
    &= \ev*{\Op{Q}{\Delta} \prod_{i=2}^{n-2} \Op{\delta_i}{q_i} \Op{\Delta'}{-Q'}}_{CFT} e^{-S_{\textsc{sb}}[\theta_{S}]}\ ,
  \end{aligned}
\end{equation}
where \(\Delta = \Delta(Q)\), \(\Delta'=\Delta(Q')\), \(Q' = Q + q_2 + \dots + q_{n-2}\), \(Q \gg q_i\), and \(\ev{\dots}_{CFT}\) is the correlator in the conformal theory.
In this limit the breaking of Schrödinger symmetry is measured by a universal term that is easily evaluated.

\begin{figure}[!ht]
\centering
\includegraphics[width = 0.7\textwidth]{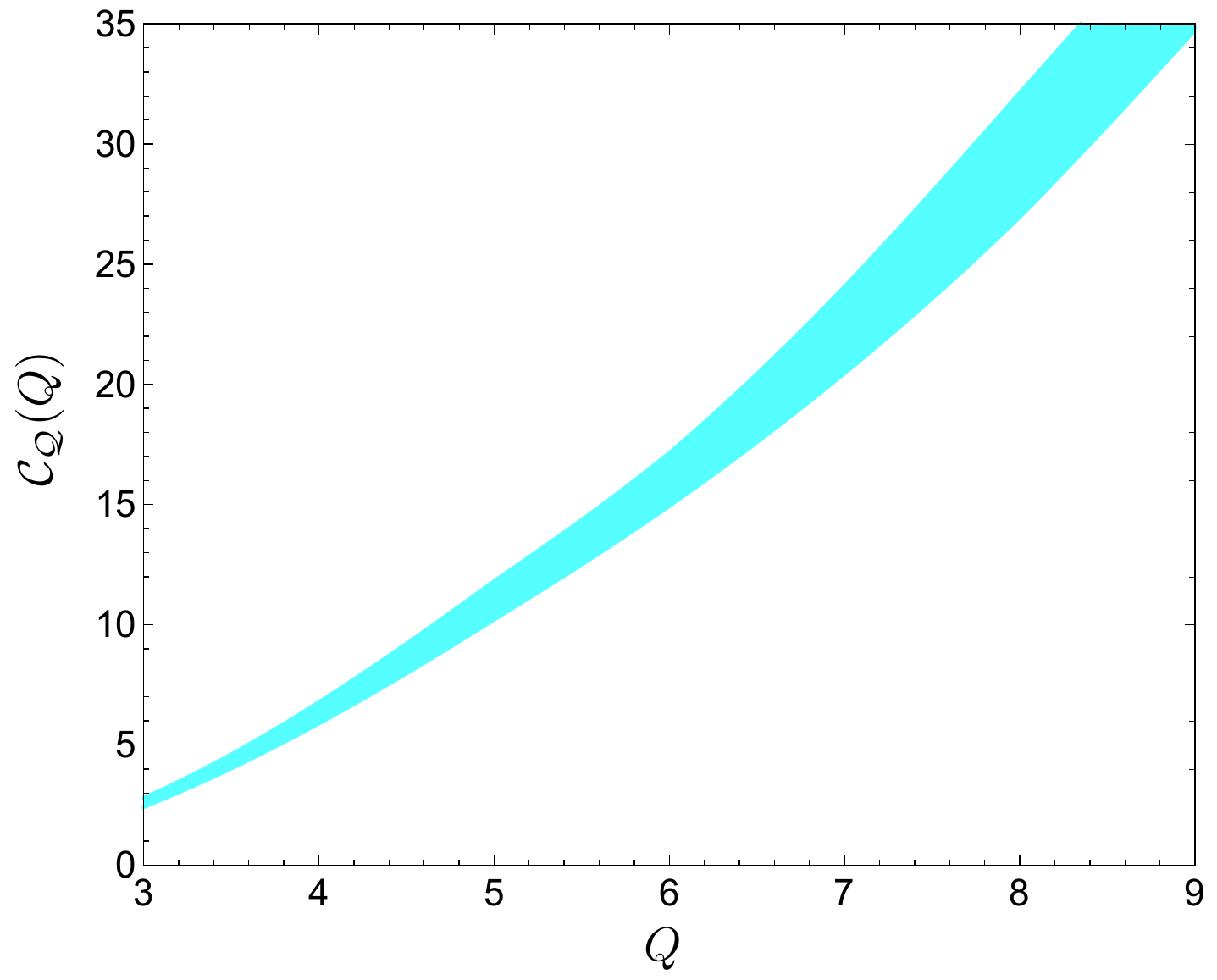}
\caption{Plot of the function ${\cal C_Q}(Q)$ vs $Q$. The band represents the uncertainty in the parameter $g_1$, determined by
Monte Carlo simulations, as discussed in the text.}
  \label{fig:CQ}
\end{figure}

For the leading scattering length correction one finds
\begin{equation}
\int d^3{\bf x}\,{\cal H}^{a}_{\textsc{sb}} \ = \ {\cal C}^{a}M^{-1/2}Q^{7/6}\,a^{-1}{\bar\omega(\tau)^{1/2}}\ ,
\label{eq:swscbe1}
\end{equation}
where
\begin{equation}
{\cal C}^{a}\equiv  \frac{64\sqrt{2}\,3^{1/6}\pi\,\xi^{7/4}g_1}{35} \ ,
\label{eq:swscbe1b}
\end{equation}   
and
\begin{eqnarray}
S^{a}_{\textsc{sb}}&=& \int \dd^4{x}\,{\cal H}^{a}_{\textsc{sb}} \ =\  {\cal C}^{a}M^{-1/2}Q^{7/6}a^{-1} \frac{1}{\sqrt{2}}\int_{\tau_2+\varepsilon}^{\tau_1-\varepsilon}d\tau\left(\frac{(\tau_1-\tau_2)}{(\tau-\tau_2)(\tau_1-\tau)}\right)^{1/2} 
\nonumber \\ & =& {\frac{1}{\sqrt{2}}\pi}\,{\cal C}^{a}M^{-1/2}Q^{7/6}a^{-1}\tau_{12}^{1/2}\ .
\label{eq:swscbe2}
\end{eqnarray}
This effect is small as compared to LO for $a^{-1} \ll M^{1/2}Q^{1/6}\tau_{12}^{-1/2}\sim \Lambda_{UV}$. Therefore, with $a$ large, one
can take $a^{-1}\sim p_{IR}$ and scattering length effects are small at large charge.
For the two-point function one finds
\begin{eqnarray}
G_Q(x_1,x_2) &=& G^{CFT}_Q(x_1,x_2)\Big\lbrack 1\ -\ {\frac{1}{\sqrt{2}}}\pi{\cal C}^{a}M^{-1/2}Q^{7/6}a^{-1}\tau_{12}^{1/2} \Big\rbrack \nonumber \ ,
\label{eq:swscbe5}
\end{eqnarray}
which is indeed of the expected form, Eq.~(\ref{eq:swscbe1a}). 
Now, continuing back to Minkowski space, choosing the source points $x_1=(t,{\bf x})$, $x_2=(0,{\bf 0})$, and using the Fourier
transform ~\cite{Hammer:2021zxb,Braaten:2023acw}
\begin{eqnarray}
&&  \int dt \int d^3{\bf x}\; \theta(t)\,t^{-\Delta}\,\exp\left({i\frac{Q M \mathbf{x}^2}{2 t}}\right) \exp\left( i E t - i{\bf p}\cdot{\bf x}\right)\nonumber \\
  && \qquad\qquad\ =\  i^{\Delta-1} \left( \frac{2\pi}{QM}\right)^{3/2}   \left(\frac{p^2}{2QM}-E\right)^{\Delta-5/2}\Gamma\left(\frac{5}{2}-\Delta\right) \ ,
\label{eq:swscbe6}
\end{eqnarray}
one finds
\begin{eqnarray}
G_Q(E,{\bf p}) &=& -i {\cal N}^2 \left( \frac{2\pi}{QM}\right)^{3/2}\Gamma\left(\frac{5}{2}-\Delta_Q\right) \left(\frac{p^2}{2QM}-E\right)^{\Delta_Q-5/2}\nonumber \\
&&\times \Bigg\lbrack 1  \ -\ {\frac{1}{\sqrt{2}}}\pi{\cal C}^{a}M^{-1/2}Q^{7/6}a^{-1} \frac{\Gamma\left(\frac{6}{2}-\Delta_Q\right)}{\Gamma\left(\frac{5}{2}-\Delta_Q\right)}\left(\frac{p^2}{2QM}-E\right)^{-1/2}
\Bigg\rbrack\;.
\label{eq:swscbe7}
\end{eqnarray}
Finally, one has
\begin{eqnarray}
{\rm Im}\;G_Q(E,{\bf 0}) &=& C_0\;E^{\Delta_Q-5/2} \Bigg\lbrack 1  \ +\ \frac{\cal C_Q}{a\sqrt{ME}}\Bigg\rbrack\; ,
\label{eq:swscbe8}
\end{eqnarray}
where $C_0$ is a normalization constant that can be absorbed into the definition of the $X$ field, and 
\begin{eqnarray}
\mkern-45mu {\cal C_Q}(Q)&=&  -{\frac{1}{\sqrt{2}}}\pi{\cal C}^{a}Q^{7/6}\frac{\Gamma\left(\frac{6}{2}-\Delta_Q\right)}{\Gamma\left(\frac{5}{2}-\Delta_Q\right)}\tan\pi\Delta_Q \ \ \ \mapright{Q\to\infty}\ \ \ 
3^{5/6}\frac{64}{175} \zeta Q^{11/6}\;+\; \mathcal{O}(Q^{1/2}),
\label{eq:swscbe9}
\end{eqnarray}
where on the right side the result at asymptotic $Q$ is given.
This function is plotted in Fig.~\ref{fig:CQ}. The $\mathcal{O}(a^{-2})$ corrections are similarly evaluated. 
It is noteworthy and promising that, for $Q\sim 3$, these $\mathcal{O}(a^{-1})$ corrections in the large-charge EFT are consistent with the range of values found in
Ref.~\cite{Chowdhury:2023ahp} working directly with the three-body wavefunctions. Of course the utility of the large-charge EFT for neutron matter requires consideration
of effective-range effects, which are sizeable in the neutron-neutron system. 

\section{Large N at large charge}
\label{sec:large-n}

The action in Eq.~\eqref{eq:lagdimerondecomp} can be studied
perturbatively without the need for an EFT if one introduces $N$
species of fermions and takes the $N \to \infty$ limit.  To make
contact with the superfluid EFT, a sector of fixed charge $Q$ is also
considered, and the double-scaling limit $Q/N$ with fixed is studied.  Here
the effect of a finite scattering length \(a\) on the gap and on the
value of the grand potential is considered.  The calculation of other
observables is left for future work.

The starting point is the action
\begin{multline}
	S  = \int\dd{t} \dd^3{\mathbf{x}} \Bigg\lbrack\sum_{i=1}^N \psi_{\sigma i}^{\dagger} \left(i \partial_t + \frac{1}{2M} \nabla^2 + \mu \right)\psi_{\sigma i} + \psi_{\downarrow i}^\dagger \psi_{\uparrow i}^\dagger  s + s^\dagger \psi_{\uparrow i} \psi_{\downarrow i} + \frac{NM}{4\pi a} s^\dagger s \\
	 -  \frac{M^2 r}{16\pi} s^\dagger\left( i\overleftrightarrow{\partial_t} + \frac{\nab^{\,2}+\nabl^{\,2}}{4M} \right) s \Bigg\rbrack\ ,
\end{multline}
which is the non-relativistic version of the \ac{gny} model.
Note that the field $s$ remains a singlet of the $U(1)\times Sp(2N)$ symmetry of the action.

One way to view this model is as the \ac{uv}
completion of a fermionic theory with four-Fermi interaction.  At high
energy, both the fermionic and the bosonic fields are dynamical.  Going
down in energy, the first scale that one meets is \(1/(M r a)\) where
the scalar field is frozen, and one is not sensitive to a finite effective
range of the interactions.  Below this energy, \(s\) is non-dynamical,
appears quadratically and can be integrated out, leading to a
four-Fermi interaction with coupling proportional to \(a/N\).  In this
way, the strongly-coupled fixed point of the fermionic model then
appears as an \ac{ir} fixed point for the non-relativistic \ac{gny}
model.  From this point of view, the theory under discussion
lives below the first scale (the bosonic field is not dynamical) and
is close enough to the \ac{ir} fixed point.

The fermions appear quadratically and can be integrated out,
\begin{equation}
  S = -N \Tr\log G^{-1}[s]  - \int \dd{\tau} \dd{\mathbf{x}} \frac{NM}{4\pi a} s^\dagger s  \ ,
\end{equation}
where the inverse propagator is given by
\begin{equation}
  G^{-1}[s] = \begin{pmatrix}
                - \partial_\tau + \frac{1}{2 M} \nabla^2 +  \mu & s(x) \\
                 s(x)^\dagger & - \partial_\tau - \frac{1}{2 M} \nabla^2 -  \mu
              \end{pmatrix}.
\end{equation}
At leading order in $N$ one can look for a saddle at constant (real) values of $s$, the effect of the fluctuations being controlled by $1/N$.
A standard calculation leads to the following expression for the action evaluated at $s=\Sigma_0$~\cite{Hellerman:2023myh}:
\begin{equation}
	\frac{S(\mu, \Sigma_0)}{N VT} = -4\pi M^{3/2} I_{0,0}\pqty*{\frac{\Sigma_0}{\mu}} \mu^{5/2} - \frac{\Sigma_0^2}{4\pi a} ,
\end{equation}
where $I_{m,n}$ is written in terms of Gaussian hypergeometric functions $\hypF$:
\begin{multline}
  I_{m,n}(y) = - \frac{y^{5/2 + m/2 - 2 n}}{2 (2 \pi)^{7/2}} \Big[ \Gamma(\tfrac{3+m}{4}) \Gamma(n - \tfrac{5+m}{4})\hypF(- \tfrac{m+1}{4}, n - \tfrac{m+5}{4}, \tfrac{1}{2} ; -\tfrac{1}{y^2 })  \\
  + \frac{2}{y}  \Gamma(\tfrac{5+m}{4}) \Gamma(n - \tfrac{3+m}{4})\hypF(\tfrac{1-m}{4}, n - \tfrac{m+3}{4}, \tfrac{3}{2} ; -\tfrac{1}{y^2 }) \Big] .
\end{multline}
Here, a heat kernel regularization has been used in which the conformal point is at $1/a^* = 0$.

The value of the action at the saddle, that is identified with the grand potential $\Omega(\mu)$, is obtained by solving the gap equation
\begin{equation}
	\frac{\dd{} }{\dd{\Sigma}}S(\mu,\Sigma) =0.
\end{equation}
This equation can be solved perturbatively in \(1/a\) using the identity
\begin{equation}
  I_{m,n}'(y) = -2 y I_{m,n+1}(y) \ .
\end{equation}
At leading order,
\begin{equation}
  I_{0,1}(y_0) = 0\ ,
\end{equation}
which can be solved numerically to give \(y_0 = \Sigma_0/\mu = 1.162\), and a Bertsch parameter of \(\xi = 0.5906 \), consistent with well-known results in the literature~\cite{Veillette:2007zz}.
Expanding around this point in inverse powers of \(a\), one then finds the gap
\begin{equation}
  \Sigma_0 = \mu y_0 \pqty*{ 1 - \frac{1}{32 \pi^2 y_0^2 a \mu^{1/2}} \frac{1}{I_{0,2}(y_0) }  + \frac{1}{2 \pqty*{32 \pi^2 y_0^2 a \mu^{1/2}}^2} \frac{1}{I_{0,2}(y_0)^2 } \dots } \ ,
\end{equation}
and the grand potential
\begin{equation}
  \begin{aligned}
    \Omega(\mu) &=
           \begin{multlined}[t]
             4\pi N \mu^{5/2} I_{0,0}(y_0) \Big(  1-\frac{y_0^2}{16 \pi^2 a \mu^{1/2}  I_{0,0}(y_0) } - \frac{1}{2 \pqty*{16 \pi^2 a \mu^{1/2}}^2 I_{0,0}(y_0) I_{0,2}(y_0)} \\
             -\frac{1}{6 \pqty*{16 \pi^2 a \mu^{1/2}}^3 y_0^2 I_{0,0}(y_0) I_{0,2}(y_0)^2 } + \dots \Big)
           \end{multlined}\\
         &= - 0.08418\  \mu^{5/2} \pqty*{ 1 + \frac{1.277}{a \mu^{1/2}} +\frac{1.513}{a^2 \mu }-\frac{1.195}{a^3 \mu ^{3/2}} + \dots} \ .
  \end{aligned}
\end{equation}
In the limit of large chemical potential, the expected
corrections to the standard result for the grand potential are found at the
fixed point, which are controlled by \(1/(a \mu^{1/2})\).  This is in
perfect analogy with the result for the \(n\)-point functions in the
previous section which, in the grand canonical ensemble, are
controlled by \({\tau_{12}^{1/2}}/(a \gamma^{1/2})\), which is consistent with
the identification of \(2 \gamma/\tau_{12}\) with the chemical
potential. The grand potential is easily read off from the Euclidean Lagrange density, including scattering length corrections, giving
\begin{equation}
    \Omega(\mu) \ =\ -c_0 \mu^{5/2} \Big(  1 \ +\ \frac{3\zeta}{2^{3/2}\xi^{1/2} a \mu^{1/2} } \ +\  \frac{3\left(4 \zeta^2+5\zeta_2 \xi\right)}{20 \xi a^2 \mu } \ +\ \ldots \Big) \ .
 \end{equation}
Matching, one finds that in the large-\(N\) limit,  \(\zeta = 0.925\) and \(\zeta_2 = 0.858\). These values are consistent with 
the Monte Carlo simulation values, if one makes a conservative estimate of uncertainties due to omitted higher orders in
the large-$N$ expansion.

\section{Conclusion}
\label{sec:conc}

Schr\"odinger symmetry places strong constraints on the form of
correlation functions in non-relativistic conformal field theory, and
the state-operator correspondence provides a powerful tool for
computing the conformal dimensions of the operators that appear in the
correlation functions. This paper has shown that in the large-charge
limit, there exists a Goldstone boson master field that allows for the
exact computation of large-charge correlation functions directly from
their path integral expressions. A direct derivation of the master
field was given which relies on the assumption of Schr\"odinger
invariance of the Goldstone-boson equation of motion. However, it has
also been shown that the master-field solution is present in the
Schr\"odinger algebra, and can be obtained directly from the
ground-state large-charge solution in the absence of sources via a
transformation from the Galilean frame to the oscillator frame.  The
master field has been used to reproduce the known $2$- and $3$-point
correlation functions, as well as the $n$-point correlation function
with an insertion of large charge.

Additionally, the grand potential has been computed including the effect of
finite scattering length in the limit of large number of fermion
species without using the EFT. The result is consistent with the EFT
results. In this limit, moreover, the numerical constants that are not
accessible in the EFT can be computed and are consistent with Monte
Carlo results.

\medskip

While Schr\"odinger symmetry constrains the form of Schr\"odinger
symmetry breaking corrections, the overall coefficients of the
corrections are not fixed by symmetry. However, in conformal
perturbation theory, these coefficients are determined by the master
field solution. In particular, symmetry-breaking corrections due to
finite scattering length were computed, and shown to appear in a
universal manner for any large-charge $n$-point function.  While
scattering length corrections are naturally accounted for in the
large-charge EFT, the effective-range corrections have to be
unnaturally small to remain a perturbative effect. This suggests that
the large-charge EFT may not provide an efficient description of
systems of many neutrons. Here the issue is subtle as in the
large-charge limit, the superfluid EFT probes all distance scales in
the fundamental-fermion EFT.  The issue of the effective range
corrections and their quantitative contribution to symmetry breaking
in neutron matter will be treated elsewhere.

It would be interesting to consider quantum corrections, which have
been studied using the state-operator correspondence, in the context
of the master field solution and the time-dependent superfluid droplet.
In addition, given the simplicity of the master-field equation, one interesting
avenue to pursue is the possibility of a holographic
dual~\cite{Son:2008ye,Balasubramanian_2008} based on a Schrödinger
background~\cite{Israel:2004vv}.

\section*{Acknowledgments}

We would like to thank Dam Thanh Son for illuminating discussions.
D.O. and S.R. thank Jan Herrmann for discussions and collaboration
related to the large-N computation, This work was supported by the Swiss National Science
Foundation (SNSF) under grant number 200021\_192137 and by the
U.~S.~Department of Energy grant {\bf DE-FG02-97ER-41014} (UW Nuclear
Theory).

\bibliographystyle{JHEP}
\bibliography{bibi}
\end{document}